\documentclass[12pt]{amsproc}
\usepackage{a4,amsmath,amssymb,hyperref,graphicx}

\allowdisplaybreaks[4]
\newtheorem{Definition}{Definition}
\newtheorem{Proposition}[Definition]{Proposition}
\newtheorem{Theorem}[Definition]{Theorem}
\newtheorem{Lemma}[Definition]{Lemma}

\title[Non-compact spectral triples with finite volume]{Non-compact
  spectral triples with finite volume} 

\author{Raimar Wulkenhaar}

\address{Mathematisches Institut der Westf\"alischen
  Wilhelms-Universit\"at, Einsteinstra\ss{}e 62, D-48149 M\"unster, Germany}

\email{raimar@math.uni-muenster.de}

\dedicatory{Dedicated to Alain Connes on the occasion of his 60th birthday}

\begin{document}

\begin{abstract}
In order to extend the spectral action principle to non-compact
  spaces, we propose a framework for spectral triples where the
  algebra may be non-unital but the resolvent of the Dirac operator
  remains compact. We show that an example is given by the 
  supersymmetric harmonic oscillator which, interestingly, provides
  two different Dirac operators. This leads to two different
  representations of the volume form in the Hilbert space, and only
  their product is the grading operator. The index of the even-to-odd part
  of each of these Dirac operators is $1$.

  We also compute the spectral action for the corresponding
  Connes-Lott two-point model. There is an additional harmonic
  oscillator potential for the Higgs field, whereas the Yang-Mills
  part is unchanged. The total Higgs potential shows a two-phase
  structure with smooth transition between them: In the spontaneously
  broken phase below a critical radius, all fields are massive, with
  the Higgs field mass slightly smaller than the NCG prediction. In the
  unbroken phase above the critical radius, gauge fields and fermions
  are massless, whereas the Higgs field remains massive.
\end{abstract}

\maketitle

\section{Introduction}

One of the greatest achievements of noncommutative geometry
\cite{Connes:1994yd} is the conceptual understanding of the Standard
Model of particle physics. This was not reached in one step. It took
more than 15 years 
\begin{itemize}
\item from the first appearance of the Higgs potential in
noncommutative models \cite{DuboisViolette:1988ps,Connes:1990??} 

\item via
the two-sheeted universe of Connes-Lott
\cite{Connes:1990qp} with its bimodule structure \cite{Connes:1994yd},

\item the discovery of the real
structure \cite{Connes:1995tu} (which eliminated one redundant $U(1)$ 
group),

\item the understanding of gauge fields as inner fluctuations in an
  axiomatic setting \cite{Connes:1996gi} and the move from the Dixmier
  trace based action functional to the spectral action principle
  \cite{Chamseddine:1996zu}, which unifies the Standard Model with
  gravity,

\item the supersession of the unimodularity condition
  \cite{Lazzarini:2001mx} (which eliminated the second redundant
  $U(1)$ group),

\item to the spectacular rebirth \cite{Chamseddine:2006ep} with the
  explanation \cite{Chamseddine:2007ia} of the $\mathbb{C}\oplus
  \mathbb{H} \oplus M_3(\mathbb{C})$ Standard Model matrix algebra as
  the distinguished maximal subalgebra of $M_2(\mathbb{H}) \oplus
  M_4(\mathbb{C})$ compatible with a non-trivial first order condition
  (i.e.\ Majorana masses) and a six-dimensional real structure
  (i.e.\ charge conjugation).
\end{itemize}

There is one important message of this evolution: One should never be
completely satisfied with one's achievements! The description given in
Alain Connes' book \cite{Connes:1994yd} definitely has its beauty. The
little annoyance with the redundant $U(1)$ found its solution in the
real structure \cite{Connes:1995tu} which soon was realised as a key
to unlocking the secrets of spin manifolds \cite{Connes:1996gi} in
noncommutative geometry. This axiomatic setting initiated many examples
of noncommutative manifolds and culminated in the recent spectral
characterisation of manifolds \cite{Connes:2008??}.

Let me give a wish list for further improvements---not as a criticism of the
model, but rather as a possible source of insight.
\begin{enumerate}
\item Quantisation. The outcome of the spectral action principle is a
  classical action functional valid at a distinguished (grand
  unification) scale. It is connected to the scale realised in a
  particle accelerator by the renormalisation group flow. This flow
  can be computed by rules from perturbative quantum field theory. The
  input is not directly the spectral action, but a gauge-fixed version
  of it which involves Faddeev-Popov ghosts. It is highly desirable to
  include these ghosts in the spectral action, because in this way
  unitary invariance is realised as cohomology of the BRS complex. We
  may speculate that the BRS cohomology of the spectral action is
  deeply connected to the wealth of noncommutative cohomology
  theories. As a starting point one might use results of Perrot
  \cite{Perrot:2006tr}, who identifies the BRS coboundary as the de Rham
  differential in the loop space
  $C^\infty(S^1,\mathcal{U}(\mathcal{A}))$ and connects the chiral
  anomaly with the local index formula \cite{Connes:1995??}.

\item Big desert. The present form of the spectral action is based on
  the big desert hypothesis which asserts that, apart from the Higgs
  boson, all particles relevant at the grand unification scale are
  already discovered. The minor mismatch between observed and
  predicted $U(1)$ coupling constant (see Figure 1 in
  \cite{Chamseddine:2006ep}) might suggest some new physics in the
  desert. Candidates include supersymmetry and dark matter, but also
  noncommutativity of space itself could alter the slope of the
  running $U(1)$ coupling. 

  The latter question concerning the renormalisation group flow of field
  theories on noncommutative geometries was intensely studied in the
  last decade. After unexpected difficulties with UV/IR-mixing, we
  established perturbative renormalisability of scalar field theories
  on Moyal-deformed Euclidean space
  \cite{Grosse:2004yu,Grosse:2005da}. The key is a deformation also of
  the differential calculus, namely from the Laplace operator to the
  harmonic oscillator Schr\"odinger operator. It turned out indeed that
  the combined Moyal-harmonic oscillator deformation removes the
  Landau ghost of the commutative scalar model \cite{Disertori:2006nq}
  by altering the slope of the running coupling constant
  \cite{Grosse:2004by}.  Since the $U(1)$-part of the Standard Model
  has the same Landau ghost problem, we might expect that, once the 
  Standard Model has been grounded in an appropriate 
  noncommutative geometry, the three
  running couplings of Figure 1 in \cite{Chamseddine:2006ep} will
  eventually intersect in a single point.

  The first step in this programme is to construct a spectral triple
  with its canonically associated spectral action for the combined
  Moyal-harmonic oscillator deformation. The present paper achieves an
  intermediate goal: We construct and investigate a \emph{commutative}
  harmonic oscillator spectral triple. Its Moyal isospectral
  deformation will be treated in \cite{Grosse:2009??}, building on
  ideas developed in \cite{Grosse:2007jy}. The main obstacle was to
  identify a Dirac operator whose square is the harmonic
  oscillator Hamiltonian of \cite{Grosse:2004yu}. The solution which
  we give in this paper is deeply connected to supersymmetric quantum
  mechanics \cite{Cooper:1994eh}, in particular to Witten's approach to
  Morse theory \cite{Witten:1982im}. It would be interesting to
  reformulate Witten's results in noncommutative index theory using the
  spectral triple we suggest.

\item Time. The spectral action relies on compact Euclidean geometry.
  For the Standard Model one typically chooses the manifold $S^3\times
  S^1$, where $S^3$ is for ``space'' and $S^1$ for ``temperature'', not
  ``time''. Although the universe is filled with thermal background
  radiation, it is desirable to allow for a genuine time evolution of
  the spectral geometry. In fact, noncommutative von Neumann algebras
  carry their own time evolution through the modular automorphism
  group, and it has been argued \cite{Connes:1994hv} that this is the
  source of the physical time flow. So far the modular automorphisms
  seem disconnected from the spectral action. The most ambitious
  project to reconcile time development and spectral geometry within
  generally covariant quantum field theory was initiated by Paschke
  and Verch \cite{Paschke:2004xf}.

\item Compactness. As mentioned above, the spectral action presumes
  compactness, namely, compactness of the resolvent of the
  Dirac operator. The example we study in this paper shows that
  compactness of the resolvent does not imply spacial
  compactness. It is eventually a matter of experiment to determine
  the type of compactness of the universe.

\end{enumerate}

The paper is organised as follows: We propose in
Section~\ref{sec:defST} a definition of non-unital spectral triples,
but with compactness of the resolvent of the Dirac operator. We show
in Section~\ref{sec:harmonic} that the supersymmetric harmonic
oscillator is an example of such a spectral triple: In
Section~\ref{sec:SUSY} we introduce the supercharges in a slightly
generalised framework and discuss briefly their cohomology. The
supercharges give rise to two distinct Dirac operators. In
Section~\ref{sec:oscillator} we identify for the harmonic oscillator
the algebra and the smooth part of the Hilbert space. In
Section~\ref{sec:dimspec} and Appendix~\ref{appendix} we compute the
dimension spectrum. The novel orientability structure is studied in
Section~\ref{sec:Orientability}, and Section~\ref{sec:index} discusses
the index formula for the Dirac operators.  The spectral action is
computed in Section~\ref{sec:spectralaction} and
Appendix~\ref{App-B}. In the final Section~\ref{sec:eq-motion} we
study the solution of the equations of motion.

\section{Non-compact spectral triples}

\label{sec:defST}

Motivated by the spectral characterisation of manifolds
\cite{Connes:2008??}, we propose here a definition of spectral triples
which does not require the algebra to be unital. There are several
proposals in the literature for a non-compact generalisation of
spectral triples, see \cite{Gayral:2003dm} and references therein. To
include the $\mathbb{R}^d$ with its standard Dirac operator, these
proposals relax the compactness of the resolvent of $\mathcal{D}$ to
the requirement that $\pi(a)(\mathcal{D}+\mathrm{i})^{-1}$ is compact
for all $a\in \mathcal{A}$. However, compactness of the resolvent (or
similar regularisation \cite{Gayral:2004ww}) is essential for a
well-defined spectral action. Moreover, the usual Dirac operator on
$\mathbb{R}^d$ is not suited for an index formula \cite{Elliott:1996??}.
We therefore keep compactness of the  resolvent
(and thus exclude standard $\mathbb{R}^d$), but to achieve this in the
non-compact situation we are forced to give up (at least in our
example)
\begin{enumerate}
\item the universality of dimensions,

\item the connection between volume form and $\mathbb{Z}_2$-grading.

\end{enumerate}
We give some comments after the definition. To simplify the
presentation we require the algebra to be commutative; the
noncommutative generalisation involves the real structure $J$.

\begin{Definition}
\label{Def:ST}
A (possibly non-compact) commutative spectral triple with finite
  volume $(\mathcal{A},\mathcal{H},\mathcal{D})$ is given by a
(possibly non-unital) commutative and involutive 
algebra $\mathcal{A}$ represented on a 
Hilbert space $\mathcal{H}$ and a selfadjoint
unbounded operator $\mathcal{D}$ in $\mathcal{H}$ with compact
  resolvent fulfilling the conditions 1-5 below.
\begin{enumerate}
\item {\tt Regularity and dimension spectrum.} \itemsep 1ex
For any $a \in \mathcal{A}$,  
both $a$ and $[\mathcal{D},a]$ belong 
to $\bigcap_{n =1}^\infty \mathrm{dom}
  (\delta^n)$, where $\delta T:=[\langle\mathcal{D}\rangle,T]$ and
  $\langle \mathcal{D}\rangle:=(\mathcal{D}^2+1)^{\frac{1}{2}}$.

  For any element $\phi$ of the algebra $\Psi_0(\mathcal{A})$
  generated by $\delta^m a$ and $\delta^m[\mathcal{D},a]$, with $a \in
  \mathcal{A}$, the function $\zeta_\phi(z):=\mathrm{Tr}(\phi
  \langle\mathcal{D}\rangle^{-z})$ extends
  holomorphically to $\mathbb{C}\setminus \mathrm{Sd}$ for some
  discrete set $\mathrm{Sd} \subset \mathbb{C}$ (the dimension
  spectrum), and all poles of $\zeta_\phi$ at $z \in \mathrm{Sd}$ are
  simple.  

\item {\tt Metric dimension.}
  The maximum $d:=\max \{ r \in \mathbb{R}\cap \mathrm{Sd}\}$ belongs
  to $\mathbb{N}$. The
  noncommutative integral $\displaystyle \int \kern -1em - \, a
  \langle \mathcal{D}\rangle^{-d}$ is finite for any $a \in
  \mathcal{A}$ and positive for positive elements of $\mathcal{A}$.

\item  {\tt Orientability.}
For the preferred unitisation
\[
\mathcal{B}:= 
\{ b \in \mathcal{A}''\;: \quad b,[\mathcal{D},b]
\in \bigcap_{n\in \mathbb{N}} \mathrm{dom} (\delta^m)\}\;,
\]
there is a Hochschild $d$-cycle $\boldsymbol{c} \in
Z_d(\mathcal{B},\mathcal{B})$, i.e.\ a finite sum
of terms $b_0 \otimes b_1 \otimes \dots \otimes b_d$. Its
representation $\boldsymbol{\gamma}:=\pi_{\mathcal{D}}( \boldsymbol{c})$, with 
$\pi_{\mathcal{D}}(b_0 \otimes b_1 \otimes \dots \otimes b_d)
:= b_0 [\mathcal{D},b_1]\cdots 
[\mathcal{D},b_d]$, satisfies $\boldsymbol{\gamma}^2=1$ and
$\boldsymbol{\gamma}^*=\boldsymbol{\gamma}$. Additionally,
$\boldsymbol{\gamma}$ defines the volume form on $\mathcal{A}$, i.e.\ 
\[
\phi_{\boldsymbol{\gamma}}(a_0,\dots,a_{d}) :=  \int \kern -1em - \,
\big(\boldsymbol{\gamma} a_0 
[\mathcal{D},a_1]\cdots  
[\mathcal{D},a_d]\langle\mathcal{D}\rangle^{-d}\big)
\]
provides a non-vanishing Hochschild $d$-cocycle
$\phi_{\boldsymbol{\gamma}}$ on $\mathcal{A}$.

\item {\tt First order.}
$[[\mathcal{D},b],b'] = 0$ for all $b,b' \in \mathcal{B}$.

\item {\tt Finiteness.}
  The subspace $\displaystyle
  \mathcal{H}_\infty:=\bigcap_{k=0}^\infty \mathrm{dom}(\mathcal{D}^k)
   \subset \mathcal{H}$ is a finitely
  generated projective $\mathcal{A}$-module $e \mathcal{A}^n$, for
  some $n \in \mathbb{N}$ and some projector $e=e^2=e^* \in
  M_n(\mathcal{B})$. The composition of the noncommutative integral
  with the induced hermitian structure $( ~|~) :
  \mathcal{H}_\infty \times \mathcal{H}_\infty \to \mathcal{A}$
  coincides with the scalar product $\langle~,~\rangle$ on
  $\mathcal{H}_\infty$,
\[
\langle\xi,\eta\rangle = \int \kern -1em - \,
\Big( (\xi|\eta) \,\langle\mathcal{D}\rangle^{-d}\Big)\;,\qquad 
\xi,\eta \in \mathcal{H}_\infty\;.
\]

\end{enumerate}
\end{Definition}
The dimension spectrum was introduced by Connes and Moscovici
\cite{Connes:1995??}  precisely to describe by a local formula the
lower-dimensional pieces in the Chern character that are ignored by
the top-dimensional Hochschild cohomology class. The local index
formula was generalised in \cite{Carey:2008??} to a larger class of
examples. We are interested in a similar situation. For non-unital
algebras we may have the characteristic values of the resolvent of
$\mathcal{D}$ run as $\mathcal{O}(n^{-\frac{1}{p}})$ for $p$ greater
than the metric dimension $d$. The dimension spectrum is the right
tool to deal with this case.

It would be interesting to know whether Definition~\ref{Def:ST},
despite its differences with Connes' original definition
\cite{Connes:2008??}, allows reconstruction of a manifold structure on
the spectrum $X=Spec(A)$ of the norm closure $A$ of $\mathcal{A}$. At
first sight, the construction of candidates for local charts only uses
the measure $\lambda$ on $X$ defined by the noncommutative integral
$\lambda (f) =\int \kern -0.9em - \,f \langle \mathcal{D}\rangle^{-d}$ for
$f \in A=C(X)$ and the fact that the Hilbert space $\mathcal{H}$ is
precisely the $L^2$-closure of $\mathcal{H}_\infty$ with respect to
$\lambda$. The details of how $\int \kern -0.9em - \, f \langle
\mathcal{D}\rangle^{-d}$ is constructed, whether as a state-independent
Dixmier trace or as a residue in the dimension spectrum, do not seem
to enter. In particular, Lemma 2.1 of \cite{Connes:2008??} holds: if
$1\in \mathcal{A}$, then $\mathcal{B}=\mathcal{A}$ (in the notation of
Definition~\ref{Def:ST}), so that conditions 3),4),5) are the same
as in \cite{Connes:2008??}, with the sole exception that
$\boldsymbol{\gamma}$ is not necessarily the $\mathbb{Z}_2$-grading
for even $d$ or $\boldsymbol{\gamma}=1$ for odd $d$. However, this
was only used for uniqueness of the noncommutative integral, which we
achieve alternatively from the dimension spectrum. 
But \cite[\S 9]{Connes:2008??} makes heavy use of the asymptotics of
the eigenvalues of $\langle \mathcal{D}\rangle^{-1}$ to prove
injectivity of the local charts; we do not know how to 
achieve this from the dimension spectrum.

\section{A spectral triple for the harmonic oscillator}

\label{sec:harmonic}

\subsection {Supersymmetric quantum mechanics}

\label{sec:SUSY}

Supersymmetric quantum mechanics provides an elegant approach to
exactly solvable quantum-mechanical models \cite{Cooper:1994eh} and
is also a powerful tool in mathematics \cite{Witten:1982im}. Our
notation is a compromise between  \cite{Cooper:1994eh} and
\cite{Witten:1982im}. 

Let $X$ be a $d$-dimensional smooth manifold with trivial cotangent
bundle and $\partial_\mu$, for $\mu=1,\dots,d$, be the basis of the
tangent space $T_xX$ induced by the coordinate functions.  On the
Hilbert space $L^2(X)$ we consider the unbounded operators
\begin{align}
a_\mu = e^{-\omega h} \partial_\mu e^{\omega h} 
= \partial_\mu + W_\mu\;,\qquad   
a_\mu^\dag = -e^{\omega h} \partial_\mu e^{-\omega h}
=-\partial_\mu + W_\mu\;,\qquad
\end{align}
where  $h$ is some real-valued function on $X$, the Morse
function \cite{Witten:1982im}, and $W_\mu(x) =\omega (\partial_\mu h)(x)$.
The resulting commutation relations are 
\begin{align}
[a_\mu,a_\nu]&=  [a^\dag_\mu,a^\dag_\nu]=0 \;,
\qquad &
[a_\mu,a_\nu^\dag]&= 2\omega \partial_\mu \partial_\nu h \;.
\end{align}
We define fermionic ladder operators $b^\mu,b^{\dag\mu}$ which
satisfy the anticommutation relations
\begin{align}
\{ b^\mu,b^\nu\}=0\;,\qquad
\{b^{\dag\mu},b^{\dag\nu}\}=0\;,\qquad
\{ b^\mu,b^{\dag\nu}\}=\delta^{\mu\nu}\;.
\label{car-d}
\end{align}
We also let all mixed commutators vanish,
$[a_\mu^{(\dag)},b^{(\dag)\nu}]=0$. 
We introduce the supercharges $\mathfrak{Q},\mathfrak{Q}^\dag$ by 
\begin{align}
\mathfrak{Q}:= a_\mu \otimes b^{\dag\mu}\;,\qquad 
\mathfrak{Q}^\dag:= a_\mu^\dag \otimes b^\mu\;.
\end{align}
Unless otherwise stated, we use Einstein's summation convention, i.e.\
summation over a pair of upper/lower greek indices from 1 to $d$ is
self-understood. The supercharges satisfy 
\begin{align}
  \{\mathfrak{Q},\mathfrak{Q}\}
=\{\mathfrak{Q}^\dag,\mathfrak{Q}^\dag\}=0\;,\quad 
\{\mathfrak{Q},\mathfrak{Q}^\dag\}=: \mathfrak{H} \;,\qquad
[\mathfrak{Q},\mathfrak{H}]=[\mathfrak{Q}^\dag,\mathfrak{H}]=0\;. 
\end{align}
The Hamiltonian $\mathfrak{H}$ introduced by the anticommutator reads
explicitly (index raising by $\delta^{\mu\nu}$)
\begin{align}
\label{Hamiltonian}
\mathfrak{H} &= \frac{1}{2}\delta^{\mu\nu}
\{a_\mu,a_\nu^\dag\} \otimes 1 + \frac{1}{2} 
[a_\mu,a_\nu^\dag]\otimes [b^{\dag\mu},b^\nu]  
\\
\nonumber
&= \big(-\partial_\mu\partial^\mu 
+ \omega^2 (\partial_\mu h)(\partial^\mu h) \big) \otimes 1 
+ \omega (\partial_\mu\partial_\nu h) 
\otimes [b^{\dag\mu},b^\nu]\;.  
\end{align}
The supercharges give rise to \emph{two} anticommuting Dirac operators 
\begin{align}
\mathcal{D}_1 &= \mathfrak{Q}+\mathfrak{Q}^\dag\;, & 
\mathcal{D}_2 &=\mathrm{i}\mathfrak{Q}-\mathrm{i} \mathfrak{Q}^\dag\;,
\\
\mathcal{D}_i^2 &= \mathfrak{H}\quad \text{for } i=1,2\;, &
\mathcal{D}_1\mathcal{D}_2+\mathcal{D}_2\mathcal{D}_1&=0\;.
\end{align}

We let $|0\rangle_f$ be the fermionic vacuum with $b^\mu
|0\rangle_f=0$. By repeated application of $b^{\dag\mu}$ one constructs
out of $|0\rangle_f $ the $2^d$-dimensional fermionic Hilbert space
$\bigwedge (\mathbb{C}^d)$ in which we label the standard orthonormal
basis as follows:
\begin{align}
| s_1,\dots,s_d\rangle_f=(b^{\dag 1})^{s_1} \dots 
(b^{\dag d})^{s_d}|0\rangle_f \;,\qquad s_\mu \in \{0,1\}\;.
\label{ssd}
\end{align}
The fermionic number operator is 
$N_f= b_\mu^{\dag} b^\mu$, with 
$N_f | s_1,\dots,s_d\rangle_f=(s_1+\dots+s_d)  |
s_1,\dots,s_d\rangle_f$. The fermionic Hilbert space is
$\mathbb{N}$-graded by $\bigwedge (\mathbb{C}^d)=\bigoplus_{p=0}^d
\Lambda^p (\mathbb{C}^d)$ with $\mathrm{dim}(\Lambda^p (\mathbb{C}^d)) 
= \binom{d}{p}$. Accordingly, the total Hilbert space
$\mathcal{H}=L^2(X) \otimes \bigwedge (\mathbb{C}^d)$ is graded by the
fermion number $\mathcal{H}=\bigoplus_{p=0}^d \mathcal{H}_p$. Note
that $\mathfrak{Q}: \mathcal{H}_p \to \mathcal{H}_{p+1}$ and 
$\mathfrak{Q}^\dag: \mathcal{H}_p \to \mathcal{H}_{p-1}$. The
induced $\mathbb{Z}_2$-grading operator is 
\begin{align}
\Gamma &= (-1)^{N_f}\;,\qquad  \Gamma^2=1\;,~\Gamma=\Gamma^*\;,\qquad
\Gamma \mathcal{D}_i+\mathcal{D}_i\Gamma=0\;.
\end{align}
Let $B_p(\omega)$ be the dimension of the $p$-th cohomology group of
$\mathfrak{Q}$, i.e.\ the number of linearly independent $\psi_p \in
\ker \mathfrak{Q} \cap \mathcal{H}_p$ that cannot be written as
$\psi_p=\mathfrak{Q} \eta_{p-1}$ for some $\eta \in
\mathcal{H}_{p-1}$.  According to Witten \cite{Witten:1982im},
$B_p(\omega)$ coincides with the Betti number $B_p$ and is deeply
connected with the Morse index $M_p$ for the function $h$: Let
$x_\alpha$ be a critical point of $h$, i.e.\ $(\partial_\mu
h)(x)=0$. If $\partial_\mu\partial_\nu h$ is regular at each of these
critical points, then $M_p$ is the number of critical points at which
$\partial_\mu\partial_\nu h$ has $p$ negative eigenvalues. The weak
Morse inequalities $M_p \geq B_p$ follow from the eigenvalue problem
for $\mathfrak{H}$ in the limit of large $\omega$.

By Hodge theory, which relies on Hilbert space structure, every
generator of the $p$-th cohomology group of $\mathfrak{Q}$ has a
unique representative $\psi$ which is also $\mathfrak{Q}^\dag$-exact
(and thus belongs to $\ker \mathfrak{H}$). Since the
$b^\mu,b^{\dag\mu}$ generate linearly independent subspaces, this means
(no summation over $\bar{\mu},\bar{\nu}$)
\begin{align}
(a_{\bar{\mu}} \otimes b^{\dag\bar{\mu}})\psi =0 \quad \text{and}  \quad 
(a_{\bar{\nu}}^\dag \otimes b^{\bar{\nu}})\psi =0 \quad \text{for all }
\bar{\mu},\bar{\nu}=1,\dots,d\;.
\end{align}
The only candidates are (up to a multiplicative constant)
\begin{align}
\psi_0 = e^{-\omega h} |0\rangle_f \quad \text{and} \quad   
\psi_d = e^{\omega h} b^{\dag 1} \dots b^{\dag d}|0\rangle_f \;.
\end{align}
For compact manifolds, where both $e^{\pm \omega h}$ are integrable,
this yields $B_0=1$ and $B_d=1$ as the only non-vanishing Betti
numbers. In the non-compact case one should choose $e^{-\omega h}$
integrable, so that $ e^{\omega h}$ is not integrable, and hence $B_p=
\delta_{p0}$. Of course, this behaviour is due to the assumption of a
trivial cotangent bundle. For more interesting topology one should
define the smooth subspace of the Hilbert space as a finitely
generated projective module.

\subsection{The harmonic oscillator}

\label{sec:oscillator}

In the following we propose a spectral triple in the sense of
Definition~\ref{Def:ST} with objects related to the harmonic
oscillator.  We will check the
axioms, but no attempt will be made to reconstruct a manifold.

\bigskip

The harmonic oscillator is obtained from the Morse function
$h=\frac{\|x\|^2}{2}=\frac{1}{2}\delta^{\mu\nu}x_\mu x_\nu$ on the
manifold $\mathbb{R}^d$. This leads to the relation
\begin{align}
[a_\mu,a_\nu^\dag]= 2\omega \delta_{\mu\nu} \;, 
\end{align}
which in turn permits a complete reconstruction of the eigenfunctions 
by repeated application of $a_\mu^\dag,b^{\dag\nu}$ to the ground state 
$\psi_0=|0\rangle_b \otimes  |0\rangle_f  \in \ker \mathfrak{H}$, with 
$
|0\rangle_b = (\tfrac{\omega}{\pi})^{\frac{d}{4}} 
e^{-\frac{\omega}{2} \|x\|^2}$.
Defining 
\begin{align}
  |n_1,\dots ,n_d\rangle_b = \frac{1}{\sqrt{n_1!\dots n_d! (2\omega)^{n_1+\dots+n_d}}} 
(a_1^\dag)^{n_1} \cdots (a_d^\dag)^{n_d} |0\rangle_b\;,\qquad
n_\mu \in \mathbb{N}\;,
\label{nnd}
\end{align}
the tensor products $|n_1,\dots ,n_d\rangle_b \otimes |s_1,\dots
,s_d\rangle_f$ of (\ref{nnd}) with (\ref{ssd}) form an orthonormal basis
of the Hilbert space $\mathcal{H}=\ell^2(\mathbb{N}^d) \otimes
\mathbb{C}^{2^d} \simeq L^2(\mathbb{R}^d) \otimes
\bigwedge(\mathbb{C}^d)$.

There are two ways of viewing the Hamiltonian (\ref{Hamiltonian}). In
the  $L^2(\mathbb{R}^d)$-representation, we have
\begin{align}
\mathfrak{H} &= H \otimes 1 
+ \omega \otimes \Sigma\;, & 
H&= -\partial_\mu\partial^\mu + \omega^2 x_\mu x^\mu\;, & 
\Sigma &= [b_\mu^\dag,b_\mu]\;,
\label{H-x}
\end{align}
i.e.\ the total Hamiltonian is the sum of the harmonic oscillator
Hamiltonian and $\omega$ times the spin matrix $\Sigma$. This
representation will be useful when considering the algebra
$\mathcal{A}$ later on which is also realised in the
$L^2(\mathbb{R}^d)$-representation. In
the $\ell^2(\mathbb{N}^d)$-representation, we have 
\begin{align}
\mathcal{D}_1^2 = \mathcal{D}_2^2 = \mathfrak{H}
= a_\mu^\dag a^\mu \otimes 1 +
2\omega \otimes b_\mu^\dag b^\mu=2\omega(N_b+N_f)\;,
\end{align}
which is up to a factor of $2\omega$ the
supersymmetric number operator:
\begin{align}
\label{EV-D2}
&\mathcal{D}_i^2 (|n_1,\dots,n_d\rangle_b \otimes   
|s_1,\dots,s_d\rangle_f) 
\\\nonumber 
&= \Big(2\omega\sum_{\mu=1}^d (n_\mu+s_\mu)\Big)
(|n_1,\dots,n_d\rangle_b \otimes   
|s_1,\dots,s_d\rangle_f)\;.
\end{align}
In particular, the kernel of $\mathcal{D}_i$ is one-dimensional, and the
resolvent of $\mathcal{D}_i$ is compact. To deal with the kernel, we
introduce
\begin{align}
  \langle \mathcal{D}\rangle:=(\mathcal{D}_1^2+1)^{\frac{1}{2}}
=(\mathcal{D}_2^2+1)^{\frac{1}{2}}
\;,\qquad 
\delta T := [\langle \mathcal{D}\rangle,T]\quad\text{for } T \in
\mathcal{B}(\mathcal{H})\;. 
\end{align}
Counting the number of eigenvalues $\leq N$ one finds that
$\langle\mathcal{D}\rangle^{-1}$ is a noncommutative infinitesimal of
order $2d$, and $\langle\mathcal{D}\rangle^{-p}$ is trace-class
for $p>2d$.  Formula (\ref{EV-D2}) also shows that
\begin{align}
\mathcal{H}_\infty :=\bigcap_{m \geq 0} \mathrm{dom}(\mathcal{D}^n) 
=\mathcal{S}(\mathbb{N}^d) \otimes \bigwedge (\mathbb{C}^d)
\simeq 
\mathcal{S}(\mathbb{R}^d) \otimes \bigwedge (\mathbb{C}^d)
\simeq 
\big(\mathcal{S}(\mathbb{R}^d) \big)^{2^d}\;,
\end{align}
which is required to be a finitely generated projective module over
the algebra of the spectral triple. We are interested here in the
commutative case so that we are led to consider the algebra
\begin{align}
\mathcal{A}=  \mathcal{S}(\mathbb{R}^d) 
\end{align}
of Schwartz class functions with standard commutative product. The
hermitian structure is pointwise the scalar product in 
$\bigwedge (\mathbb{C}^d)$, i.e.\ 
$(\xi|\eta)=\sum_{i=1}^{2^d} \xi_i^* \eta_i$ for
$\xi=(\xi_1,\dots,\xi_{2^d}),\eta=(\eta_1,\dots,\eta_{2^d})
 \in \mathcal{H}_\infty 
= \big(\mathcal{S}(\mathbb{R}^d) \big)^{2^d}$.

As usual, we represent the algebra $\mathcal{A}$ on $\mathcal{H}$ by pointwise
multiplication in $L^2(\mathbb{R}^d)$:
\begin{align}
f (\psi \otimes \rho) :=  (f \psi) \otimes \rho\qquad 
\text{for } f\in \mathcal{A}\;,~
\psi \in L^2(\mathbb{R}^d)\;,~
\rho \in \bigwedge(\mathbb{C}^d)\;.
\end{align}
The action of $\mathcal{A}$ commutes with $b^\mu,b^{\dag\mu}$ so that
we obtain
\begin{align}
[\mathcal{D}_1,f]&= \partial_\mu f \otimes (b^{\dag\mu}-b^\mu)\;,&
[\mathcal{D}_2,f]&= \partial_\mu f \otimes (\mathrm{i}b^{\dag\mu}+
\mathrm{i} b^\mu)\;.
\label{Df}
\end{align}
In particular,
the first-order condition is satisfied.
For $f \in \mathcal{A}$, the expansion coefficients $\langle
n_1,\dots,n_d|f|n_1',\dots,n_d'\rangle$ are Schwartz sequences in
$n_\mu,n_\mu'$. Therefore, $f$ and $[\mathcal{D}_i,f]$ belong for any
$m \in \mathbb{N}$ to the domain of $\delta^m$.

We show in joint work with H.~Grosse \cite{Grosse:2009??}
(which supersedes \cite{Grosse:2007jy}), that
the Moyal-deformation of $\mathcal{S}(\mathbb{R}^d)$ together with the
same Dirac operator and Hilbert space forms a noncommutative spectral
triple in the sense of Definition~\ref{Def:ST}, i.e.\ an isospectral
deformation.

\subsection{Dimension spectrum}

\label{sec:dimspec}

In this subsection we take for $\mathcal{D}$ either of $\mathcal{D}_1$ or 
$\mathcal{D}_2$. We consider the algebra
$\Psi_0(\mathcal{A})$ generated by $\delta^mf$ and
$\delta^m[\mathcal{D},f]$. As $\langle \mathcal{D}\rangle^{-z}$ is
trace-class for $\mathrm{Re}(z)>2d$, the $\zeta$-function
$\zeta_\phi(z):=\mathrm{Tr}(\phi \langle \mathcal{D}\rangle^{-z})$
exists for such $z \in \mathbb{C}$ and $\phi \in \Psi_0(\mathcal{A})$
and can possibly be extended to a meromorphic function on
$\mathbb{C}$. The following theorem identifies the poles and the
structure of the residues:

\begin{Theorem}
\label{thm-spectrum}
  The spectral triple $(\mathcal{A},\mathcal{H},\mathcal{D})$ 
  has dimension spectrum $\mathrm{Sd}=d-\mathbb{N}$ and hence metric
  dimension $d$. All poles of $\zeta_\phi$ at $z \in
  \mathrm{Sd}$ are simple with local residues, 
i.e.\ for $\phi=\delta^{n_1}f_1\cdots
  \delta^{n_v}f_v$, any residue 
$\mathrm{res}_{z \in \mathrm{Sd}} \zeta_\phi(z)$ 
is a finite sum of
$\displaystyle \int_{\mathbb{R}^d} dx\;x^{\alpha_0} 
(\partial^{\alpha_1}f_1)\cdots
(\partial^{\alpha_v}f_v)$, where $\alpha_i$ are multi-indices. 
The analogous result holds when $f_i$ in $\phi$ is 
replaced by $[\mathcal{D},f_i]$. 
\end{Theorem}
This theorem is the central result of this paper. 
We give the rather long proof in  Appendix \ref{appendix}.

\bigskip

A special case of the proof of Theorem~\ref{thm-spectrum} is the
computation of the Dixmier trace: 
\begin{Proposition}
$\displaystyle \int \kern -1em - \;  f \langle \mathcal{D}\rangle^{-d}
=\frac{1}{(4\pi)^{\frac{d}{2}} \Gamma(\frac{d+2}{2})} 
\int_{\mathbb{R}^d} dx\; f(x)$ \quad for any $f \in \mathcal{A}$.
\end{Proposition}
\emph{Proof.} As the dimension spectrum is simple,
the Dixmier trace can be computed as a residue \cite{Carey:2003??}, 
is independent of the state $\omega$, and defines unambiguously 
the noncommutative integral: 
\begin{align}
\int \kern -1em - \; 
f \langle \mathcal{D}\rangle^{-d} = \mathrm{res}_{s=1}
\mathrm{Tr}(f \langle \mathcal{D}\rangle^{-sd})\;.
\end{align}
Taking $v=1$ and $n_1=0$ in (\ref{zeta}) and inserting $\det Q$ and
$Q^{-1}$ from (\ref{detQ}) and (\ref{invQ}) as well as
(\ref{trSigma}), we have
\begin{align}
\int \kern -1em - \; f \langle \mathcal{D}\rangle^{-d} &
= \mathrm{res}_{s=1}
\bigg( \frac{1}{\Gamma(\frac{sd}{2})}
\int_0^\infty dt_0 \;t_0^{\frac{sd}{2}-1} \,e^{-t_0}
\int_{\mathbb{R}^d} \frac{dp}{(2\pi)^d} \;\hat{f}(p) 
\frac{e^{-\frac{p^2}{\omega \tanh(\omega t_0)}}}{
\tanh^d (\omega t_0)}
\bigg)\;.
\end{align}
We write 
$\displaystyle \hat{f}(p)=\hat{f}(0) 
+p_\mu \frac{\partial \hat{f}}{\partial p_\mu}(0)
+p_\mu p_\nu \int_0^1 d\lambda\;(1-\lambda)
\frac{\partial^2 \hat{f}}{\partial p_\mu \partial p_\nu}(\lambda p_\mu)
$ and get 
\begin{align}
&
\frac{1}{\Gamma(\frac{sd}{2})}
\int_0^\infty dt_0 \;t_0^{\frac{sd}{2}-1} e^{-t_0}
\int_{\mathbb{R}^d} \frac{dp}{(2\pi)^d} \;\hat{f}(0) 
\frac{e^{-\frac{p^2}{\omega \tanh(\omega t_0)}}}{
\tanh^d (\omega t_0)}
\\
&
= \frac{\hat{f}(0)}{(4\pi)^{\frac{d}{2}}\Gamma(\frac{sd}{2})}
\int_0^\infty dt_0 \;t_0^{\frac{(s-1)d}{2}-1}\,e^{-t_0}\,
\underbrace{\Big(\frac{\omega t_0}{
\tanh (\omega t_0)}\Big)^{\frac{d}{2}}}_{g(t_0)}
\nonumber
\\
&= \frac{\hat{f}(0)}{(4\pi)^{\frac{d}{2}}}
\frac{\Gamma(\frac{(s-1)d}{2})}{\Gamma(\frac{sd}{2})}
+  \frac{\hat{f}(0)}{(4\pi)^{\frac{d}{2}}\Gamma(\frac{sd}{2})}
\int_0^\infty dt_0 \;t_0^{\frac{(s-1)d}{2}}\,e^{-t_0}\,
\int_0^1 d\lambda \;g'(\lambda t_0)\;.
\nonumber
\end{align}
As $|g'(y)|  \leq \frac{d}{2} y^{\frac{d}{2}-1}$ for all $y \in
\mathbb{R}_+$, we have
\begin{align}
\Big| \int_0^\infty dt_0 \;t_0^{\frac{(s-1)d}{2}}\,e^{-t_0}\,
\int_0^1 d\lambda \;g'(\lambda t_0)\Big| 
\leq   \int_0^\infty dt_0 \;t_0^{\frac{s}{2}-1}\,e^{-t_0}
=\Gamma(\tfrac{s}{2})\;,
\end{align}
which is regular for $s=1$. The first-order term 
$p_\mu \frac{\partial \hat{f}}{\partial p_\mu}(0)$ does not contribute
as an odd function in $p$. In the remainder, $\int_0^1 d\lambda (1-\lambda)
\frac{\partial^2 \hat{f}}{\partial p_\mu \partial_\nu}(\lambda p_\mu)$ 
is bounded, and
\begin{align}
\int \frac{dp}{(2\pi)^d} \;p_\mu p_\nu
\frac{e^{-\frac{p^2}{\omega \tanh(\omega t_0)}}}{
\tanh^d (\omega t_0)}
=\frac{\omega^2}{2} \frac{\delta_{\mu\nu}}{(4\pi)^{\frac{d}{2}}} 
\Big(\frac{\omega}{\tanh (\omega t_0)}\Big)^{\frac{d}{2}-1}
\end{align}
provides another factor of $t_0$ so that the remainder 
does not contribute to the residue at $s=1$. The assertion follows from 
$\hat{f}(0)=\displaystyle \int_{\mathbb{R}^d} dx \;f(x)$. \hfill $\square$%

\bigskip

\noindent
Therefore, with the normalisation 
$\displaystyle 
\langle\xi,\eta\rangle=
\frac{1}{(4\pi)^{\frac{d}{2}} \Gamma(\frac{d+2}{2})} 
\int_{\mathbb{R}^d} dx\; (\xi|\eta)$
of the scalar product in $\mathcal{H}$, the finiteness condition is
satisfied. 

It remains to discuss the orientability, for which we need the algebra 
\begin{align}
\mathcal{B}:= \{ b \in \mathcal{A}''\;: \quad b,[\mathcal{D},b]
\in \bigcap_{m\in \mathbb{N}} \mathrm{dom} (\delta^m)\}\;.
\end{align}
Clearly, $\mathcal{B}$ is unital and commutative;  we now show
that it contains the plane waves $u_\mu=e^{ix_\mu}$.
\begin{Lemma}
$u_\mu=e^{ix_\mu} \in \mathcal{B}\;.$
\end{Lemma}
\emph{Proof.} From (\ref{deltaT}), which applies without change to
  $T=u_\mu$, we get (no summation over $\mu$)
\begin{align}
\delta^n u_\mu &=
\frac{(-\mathrm{i})^n}{\pi^n} \int_0^\infty \prod_{i=1}^n
\frac{d\lambda_i\,\sqrt{\lambda_i}}{\langle \mathcal{D}\rangle^2 
+ \lambda_i} 
\{\underbrace{\partial_{\mu}, \dots ,\{ \partial_{\mu}}_{n\text{ 
derivatives}}, e^{\mathrm{i} x^\mu}
\} \dots \}
\prod_{j=1}^n \frac{1}{\langle \mathcal{D}\rangle^2 + \lambda_j}\;.
\end{align}
We have
\begin{align}
&\Big(\prod_{i=1}^n \frac{1}{A+\lambda_i} \Big) B 
\\
&= \bigg(\sum_{S\in \{1,2,\dots,n\}}(-1)^{|S|}
\Big(\prod_{i \in S}\frac{1}{A+\lambda_i} \Big)
(\mathrm{ad}(A))^{|S|}(B)\bigg)
\Big(\prod_{j=1}^n \frac{1}{A+\lambda_j} \Big)\;,
\nonumber
\end{align}
where the sum runs over all subsets $S \subset \{1,2,\dots,n\}$ including
the empty set. After relabelling of the $|S|$ elements of $S$, which
gives a factor $\binom{n}{|S|}$, we have  
\begin{align}
\delta^n (u_\mu) 
&= \frac{(-\mathrm{i})^n}{\pi^n}\sum_{k=0}^n \binom{n}{k} \mathrm{i}^k 
\\* \nonumber
&\times \int_0^\infty 
\prod_{i =1}^k 
\frac{1}{\langle\mathcal{D}\rangle^2 +  \lambda_i}  
\{\underbrace{\partial_{\mu}, \dots ,\{ \partial_{\mu}}_{n+k\text{ 
derivatives}}, e^{\mathrm{i} x^\mu}
\} \dots \}
\prod_{j=1}^n \frac{d\lambda_j\,\sqrt{\lambda_j}}{
(\langle\mathcal{D}\rangle^2 + \lambda_j)^2}  \;.
\end{align}
The anticommutators can be arranged as a finite sum with 
$r\leq n$ derivatives 
on the right and $l\leq k$ derivatives on the left of 
$e^{\mathrm{i} x^\mu}$. Each such term is estimated by
\begin{align}
&\bigg\|  
\int_0^\infty 
\prod_{i =1}^k 
\frac{1}{\langle\mathcal{D}\rangle^2 +  \lambda_i}  
(\partial_{\mu})^l e^{\mathrm{i} x^\mu}
(\partial_{\mu})^r \langle\mathcal{D}\rangle^{-n}
\prod_{j=1}^n \frac{d\lambda_j\,\sqrt{\lambda_j}\langle\mathcal{D}\rangle}{
(\langle\mathcal{D}\rangle^2 + \lambda_j)^2} \bigg\|
\\*
\nonumber
& \leq \big\|\langle\mathcal{D}\rangle^{-2k}(\partial_{\mu})^l \big\|\;
\big\|(\partial_{\mu})^r \langle\mathcal{D}\rangle^{-n}\big\|
\bigg\|  
\int_0^\infty 
\frac{d\lambda\,\sqrt{\lambda} \langle\mathcal{D}\rangle}{
(\langle\mathcal{D}\rangle^2 + \lambda)^{2}} \bigg\|^n\;,
\end{align}
which is bounded because the integral in the second line evaluates to
$\frac{\pi}{2}$. \hfill $\square$%

\bigskip

By the same arguments one shows that the algebra
$C_b^\infty(\mathbb{R}^d)$ of smooth bounded
functions with all derivatives bounded is contained in $\mathcal{B}$,
and it is plausible that actually 
$\mathcal{B}=C_b^\infty(\mathbb{R}^d)$.

\subsection{Orientability}

\label{sec:Orientability}

Here the distinction between $\mathcal{D}_1$ and $\mathcal{D}_2$ is
crucial again. It follows from the standard example of the compact
case that
\begin{align}
\boldsymbol{c} = \sum_{\sigma \in S_d} \epsilon(\sigma)
\frac{\mathrm{i}^{\frac{d(d-1)}{2}}}{d!} 
(u_1\cdots u_d)^{-1} \otimes u_{\sigma(1)} \otimes \dots 
u_{\sigma(d)} \in Z_d(\mathcal{B},\mathcal{B}) 
\end{align}
is a Hochschild $d$-cycle, $b\boldsymbol{c}=0$. From (\ref{Df}) and
(\ref{car-d}) we obtain
\begin{align}
\boldsymbol{\gamma}_1 
&:= \pi_{\mathcal{D}_1}(\boldsymbol{c}) = \mathrm{i}^{\frac{d(d+1)}{2}} 
(b^{\dag 1} - b^1)\cdots   (b^{\dag d} - b^d)\;, 
\\ \nonumber
\boldsymbol{\gamma}_2 &:= 
\pi_{\mathcal{D}_2}(\boldsymbol{c}) = \mathrm{i}^{\frac{d(d+3)}{2}} 
(b^{\dag 1} + b^1)\cdots   (b^{\dag d} + b^d)\;. 
\end{align}
Both $\boldsymbol{\gamma}_i$ commute with every element of
$\mathcal{A}$ or $\mathcal{B}$. 
Using the anticommutation relations (\ref{car-d}) and $(b^\mu)^* \equiv 
b^{\dag\mu}$, we have
\begin{align}
\boldsymbol{\gamma}_1^2=1=
\boldsymbol{\gamma}_2^2 \;,\qquad
\boldsymbol{\gamma}_1^*=\boldsymbol{\gamma}_1\;,\quad
\boldsymbol{\gamma}_2^*=\boldsymbol{\gamma}_2\;.
\end{align}
Decomposing the fermionic part of the Dirac operators $\mathcal{D}_i$
in $b^{\dag\mu} \pm b^\mu$, we have
\begin{align}
(b^{\dag\mu} \pm b^\mu) \boldsymbol{\gamma}_1 &= \pm(-1)^d 
\boldsymbol{\gamma}_1 (b^{\dag\mu} \pm b^\mu) \;, &
(b^{\dag\mu} \pm b^\mu) \boldsymbol{\gamma}_2 &= \mp (-1)^d 
\boldsymbol{\gamma}_2 (b^{\dag\mu} \pm b^\mu) \;.
\end{align}
Therefore, $b^{\dag\mu} \pm b^\mu$ and hence $\mathcal{D}_i$
\emph{always} ($d$ even or odd) anticommute with the product
$\boldsymbol{\gamma}_1 \boldsymbol{\gamma}_2$, which turns out to be
(up to a factor) the $\mathbb{Z}_2$-grading $(-1)^{N_f}$ of the
Hilbert space:
\begin{align}
(-\mathrm{i})^d  \boldsymbol{\gamma}_1 \boldsymbol{\gamma}_2 
= \mathrm{i}^d   \boldsymbol{\gamma}_2 \boldsymbol{\gamma}_1 
= 
(b^1 b^{\dag 1}-b^{\dag 1} b^1) \cdots 
(b^d b^{\dag d}-b^{\dag d} b^d) =(-1)^{N_f}\;.
\end{align}
This is quite different from conventional spectral triples
\cite{Connes:2008??} with a single operator $\mathcal{D}$.

\subsection{The index formula}

\label{sec:index}

We let $\mathcal{H}=\mathcal{H}_{ev} \oplus \mathcal{H}_{odd}$ be the
decomposition into even and odd subspaces with respect to the fermion
number operator $(-1)^{N_f}$. The $\mathcal{D}_i$ are off-diagonal in
this decomposition, $\mathcal{D}_i=\mathcal{D}_i^+ + \mathcal{D}_i^-$,
with $\mathcal{D}_i^+ =\mathcal{D}_i\big|_{\mathcal{H}_{ev}} :  
\mathcal{H}_{ev}\to  \mathcal{H}_{odd}$ and 
$\mathcal{D}_i^- =(\mathcal{D}_i^+)^*  
=\mathcal{D}_i\big|_{\mathcal{H}_{odd}} :  
\mathcal{H}_{odd}\to  \mathcal{H}_{ev}$. 

There is a well-defined index problem for $\mathcal{D}_i^+$ due to
Elliott, Natsume and Nest \cite{Elliott:1996??}. The $\mathcal{D}_i^+$
are elliptic  pseudodifferential operators in the sense of Shubin
\cite{Shubin:1987??} with symbol $\mathfrak{a}_i$. Then, the analytic index 
\begin{align}
\mathrm{index}\,(\mathcal{D}_i^+) = \mathrm{dim}\,\ker \mathcal{D}_i^+ 
- \mathrm{dim}\,\ker \mathcal{D}_i^- 
\end{align}
can be computed by an index formula for the symbol
$\mathfrak{a}_i$ as described below.

Following \cite{Elliott:1996??}, we associate to (appropriate)
operators $\mathcal{P}_{\mathfrak{a}}:
\mathcal{S}(\mathbb{R}^n; \mathbb{C}^k) \to 
\mathcal{S}(\mathbb{R}^n; \mathbb{C}^k)$ the symbol
symbol $\mathfrak{a} \in M_k(C^\infty(T^*\mathbb{R}^n))$
by
\begin{align}
(\mathcal{P}_{\mathfrak{a}} \eta)(x) = \frac{1}{(2\pi)^n} 
\int_{\mathbb{R}^n\times \mathbb{R}^n} d\xi \,dy
\; e^{\mathrm{i}\langle x-y,\xi\rangle} \, \mathfrak{a}_i(x,\xi)\,
\eta(y) \;,\qquad 
\eta \in \mathcal{S}(\mathbb{R}^n; \mathbb{C}^k) \;.
\end{align}
The symbol $\mathfrak{a}$ is said to be 
\emph{elliptic of order $m$} if there exist $C,R>0$ such that 
$\mathfrak{a}(x,\xi)^*\mathfrak{a}(x,\xi) \geq C (\|x\|^2+\|\xi\|^2)^m 1_k$
for $\|x\|^2+\|\xi\|^2 \geq R$. 

For $m>0$ one defines the graph projector 
\begin{align}
e_{\mathfrak{a}} =   
\left( \begin{array}{cc} 
(1+\mathfrak{a}^*\mathfrak{a})^{-1} &
(1+\mathfrak{a}^*\mathfrak{a})^{-1} \mathfrak{a} \\ 
\mathfrak{a}^* (1+\mathfrak{a}^*\mathfrak{a})^{-1} &
\mathfrak{a}^* (1+\mathfrak{a}^*\mathfrak{a})^{-1} \mathfrak{a} 
\end{array}\right) \in M_{2k}(C(T^*\mathbb{R}^n))
\end{align}
and the matrix $\hat{e}_{\mathfrak{a}} = 
e_{\mathfrak{a}} - \left(\begin{array}{cc} 0 & 0 \\ 0 & 1
  \end{array}\right) \in M_{2k}(C_0(T^*\mathbb{R}^n))$, i.e.\ 
$\hat{e}_{\mathfrak{a}}$ vanishes at infinity for $m>0$ (the entries
of $\hat{e}_{\mathfrak{a}}$ are of order $-m$). Using continuous
fields of $C^*$-algebras, the following index theorem is proven in
\cite{Elliott:1996??}:
\begin{Theorem}
If $\mathcal{P}_{\mathfrak{a}}$ is an elliptic pseudodifferential
operator of positive order, then
\begin{align}
\mathrm{index}\,(  \mathcal{P}_{\mathfrak{a}})
=\frac{1}{(2\pi\mathrm{i})^n n! }
\int_{T^*\mathbb{R}^n} \mathrm{tr}\big(  \hat{e}_{\mathfrak{a}}
(d \hat{e}_{\mathfrak{a}})^{2n}\big)\;,
\end{align}
where $T^*\mathbb{R}^n$ is oriented by 
$dx_1 \wedge d\xi_1 \wedge\dots\wedge 
dx_n \wedge d\xi_n>0$.
\end{Theorem}

Back to our example. Restricting $\mathcal{D}_i^+$ to the even part of 
$\mathcal{H}_\infty$, we regard $\mathcal{D}_i^+:
\mathcal{S}(\mathbb{R}^d; \mathbb{C}^{2^{d-1}}) \to 
\mathcal{S}(\mathbb{R}^d; \mathbb{C}^{2^{d-1}})$. The symbol 
$\mathfrak{a}_i \in  M_{2^{d-1}}(C^\infty(T^*\mathbb{R}^d))$ of
$\mathcal{D}_i^+$ is
obtained from the action of $\mathfrak{Q},\mathfrak{Q}^\dag$ 
on the basis $\mathrm{e}^{\mathrm{i}\langle \xi,x\rangle} 
|s_1,\dots,s_d\rangle_f$. 
For example, we have for $d=2$ in the matrix bases 
$\binom{|0,0\rangle_f}{|1,1\rangle_f}$ of 
$\big(\bigwedge(\mathbb{C}^d)\big)_{ev}$ and 
$\binom{|1,0\rangle_f}{|0,1\rangle_f}$ of 
$\big(\bigwedge(\mathbb{C}^d)\big)_{odd}$ the representation
\begin{align}
\mathfrak{a}_1(x_1,x_2,\xi_1,\xi_2)  
=\left( \begin{array}{cc}
\mathrm{i}\xi_1 + \omega x_1 & -(-\mathrm{i}\xi_2 +\omega  x_2) \\
\mathrm{i}\xi_2 +\omega  x_2 & -\mathrm{i}\xi_1 +\omega  x_1
\end{array}\right)\;.
\end{align}

The product $\mathfrak{a}_i(x,\xi)^*\mathfrak{a}_i(x,\xi)$ is the
restriction of the symbol of $H$ to the even subspace.
This implies 
\begin{align}
\mathfrak{a}_i(x,\xi)^*\mathfrak{a}_i(x,\xi) 
=  (\omega^2 \|x\|^2+\|\xi\|^2) 1_{2^{d-1}}\;,
\end{align}
i.e.\ ellipticity of order $1$ if $\omega>0$. Note that the usual
Dirac operator $\mathrm{i} \gamma^\mu \partial_\mu$ on $\mathbb{R}^d$
is not elliptic in this sense.

For $d=2$ an already lengthy computation shows 
\begin{align}
\mathrm{tr}\big(  \hat{e}_{\mathfrak{a}_1}\,
d \hat{e}_{\mathfrak{a}_1}\wedge 
d \hat{e}_{\mathfrak{a}_1}\wedge 
d \hat{e}_{\mathfrak{a}_1}\wedge 
d \hat{e}_{\mathfrak{a}_1}\big)=
-\frac{96\omega^2\, dx_1\wedge d\xi_1\wedge dx_2\wedge d\xi_2
}{(1+\omega^2 x_1^2+\omega^2 x_2^2+\xi_1^2+\xi_2^2)^5}
\;,
\end{align}
which yields
\begin{align}
\mathrm{index}\,(\mathcal{D}_1^+) 
= \frac{1}{(2\pi \mathrm{i})^2 \cdot 2} 
\int_0^\infty 
2\pi x\,dx \int_0^\infty 2\pi\xi\,d\xi\;
 \frac{(-96 \omega^2)}{(1+\omega^2 x^2 +\xi^2)^5}
=1\;.
\end{align}
This is of course expected in any dimension $d$: the (one-dimensional)
kernel of $\mathcal{D}_i^+$ is spanned by the Gau\ss{}ian
$e^{-\frac{\omega}{2}\|x\|^2}|0,\dots,0\rangle_f$, and the cokernel is trivial.

\section{The spectral action for the $U(1)$-Higgs model }

\label{sec:spectralaction}

In the Connes-Lott spirit \cite{Connes:1990qp} we take the tensor
product of the ($d=4$)-dimensional spectral triple
$(\mathcal{A},\mathcal{H},\mathcal{D}_1)$ with the finite
Higgs spectral triple $(\mathbb{C}\oplus\mathbb{C}, \mathbb{C}^2,
M\sigma_1,\sigma_3)$, which is even with $\mathbb{Z}_2$-grading $\sigma_3$. 
Here, $M$ is a real number, and $\sigma_k$ are the Pauli matrices. 
For the bosonic sector considered here
only the spectrum of $\mathcal{D}_i$ matters, so that 
$\mathcal{D}_1$ and $\mathcal{D}_2$ give identical results. 
The total Dirac operator $\mathcal{D}=\mathcal{D}_1 \otimes \sigma_3 +
1\otimes M\sigma_1$ of the product triple becomes
\begin{align}
\mathcal{D} =\left(\begin{array}{cc} 
\mathcal{D}_1 & M  \\ 
M  & -\mathcal{D}_1
\end{array}\right)\;.
\label{Dirac-CL}
\end{align}
In this representation, the algebra is $\mathcal{A} \oplus \mathcal{A}
\ni (f,g)$ with diagonal action by pointwise multiplication on 
$\mathcal{H}_{tot}=\mathcal{H} \oplus \mathcal{H}$.
The commutator of $\mathcal{D}$ with $(f,g)$ is
\begin{align}
[\mathcal{D},(f,g)] =\left(\begin{array}{cc} 
\partial_\mu f \otimes (b^{\dag\mu} -b^\mu) & M(g-f) \\
M (f-g) & - \partial_\mu g \otimes (b^{\dag\mu} -b^\mu)
\end{array}\right)\;.
\end{align}
This shows that selfadjoint fluctuated Dirac operators $\mathcal{D}_A=
\mathcal{D} + \sum_i a_i [\mathcal{D},b_i]$ are of the
form 
\begin{align}
\mathcal{D}_A  =\left(\begin{array}{cc} 
\mathcal{D}_1 + \mathrm{i} A_\mu \otimes  (b^{\dag\mu} -b^\mu) 
  & \phi \otimes 1\\ \bar{\phi} \otimes 1 & 
-\mathcal{D}_1 - \mathrm{i} B_\mu \otimes  (b^{\dag\mu} -b^\mu) 
\end{array}\right)\;,
\end{align}
for real fields $A_\mu=\overline{A_\mu} ,\; B_\mu=\overline{B_\mu} 
\in \mathcal{A}$ and a complex field 
$\phi \in \mathcal{A}$. The square of $\mathcal{D}_A$ is
\begin{align}
\mathcal{D}_A^2 &= \left(\begin{array}{cc} 
H\otimes 1+ \omega \otimes \Sigma +\mathrm{i}F_A + |\phi|^2 \otimes 1  
& D_\mu \phi 
\otimes  (b^{\dag\mu} -b^\mu) 
\\
-\overline{D_\mu \phi} \otimes  (b^{\dag\mu} -b^\mu) 
&
H\otimes 1+ \omega\otimes \Sigma +\mathrm{i}F_B + |\phi|^2 \otimes 1  
\end{array}\right)\;,
\label{DA2}
\end{align}
where 
\begin{align} 
D_\mu \phi &:= \partial_\mu \phi + \mathrm{i}(A_\mu-B_\mu) \phi\;,
\label{Dphi}
\\[1ex]
F_A &:= \{\mathcal{D}_1, A_\mu \otimes (b^{\dag\mu} -b^\mu)\}
+\mathrm{i} A_\mu A_\nu \otimes (b^{\dag\mu} -b^\mu)(b^{\dag\nu} -b^\nu)
\nonumber
\\
&= (-\{\partial_\mu,A^\mu\}-\mathrm{i} A_\mu A^\mu ) \otimes 1
+ \frac{1}{4} F^A_{\mu\nu} \otimes [b^{\dag\mu}-b^\mu,b^{\dag\nu}
-b^\nu]
\label{FA}
\end{align}
and similarly for $F_B$. Here, $F^A_{\mu\nu}=\partial_\mu
A_\nu-\partial_\nu A_\mu$ is the $U(1)$-curvature (field strength),
and the explicit appearance of $x$ has dropped in $F_A$ because of
$\{b^{\dag\mu}+b^\mu,b^{\dag\nu}-b^\nu\}=0$. 

\bigskip

According to the spectral action principle
\cite{Connes:1996gi,Chamseddine:1996zu}, the bosonic action depends
only on the spectrum of the Dirac operator. Thus, by functional
calculus, the most general form of the bosonic action is
\begin{align}
S(\mathcal{D}_A) &=\mathrm{Tr}\big(\chi(\mathcal{D}_A^2)\big)
=\int_0^\infty dt\; 
\mathrm{Tr}(e^{-t \mathcal{D}_A^2}) \hat{\chi}(t)\;,  
\label{SpecAc}
\end{align}
for some function $\chi:\mathbb{R}_+ \to \mathbb{R}_+$ for which the
operator trace exists. The second equality is obtained by Laplace 
transformation, which produces the inverse Laplace transform 
$\hat{\chi}$ of $\chi(s)=\int_0^\infty dt\;e^{-st} \hat{\chi}(t)$. 
One has
\begin{align}
\chi_{z} &:= \int_0^\infty  dt\;t^z \hat{\chi}(t)
= \left\{ \begin{array}{cl}\displaystyle
\frac{1}{\Gamma(-z)}\int_0^\infty
ds\;s^{-z-1} \chi(s)  & \text{for } z \notin \mathbb{N}\;,
\\[2ex]
(-1)^{k}  \chi^{(k)}(0) & \text{for } z=k \in \mathbb{N} \;.
\end{array}\right.
\end{align}

To compute the traces $\mathrm{Tr}(e^{-t \mathcal{D}_A^2})$ we write 
$\mathcal{D}_A^2=\mathrm{H}_0-V$, with $\mathrm{H}_0
:=H+\omega\Sigma$, and consider the Duhamel expansion
\begin{align}
&e^{-t_0 (\mathrm{H}_0-V)} 
\label{Duhamel}
\\
&=
 e^{-t_0 \mathrm{H}_0} - \int_0^{t_0}  \!\!dt_1 \;
\frac{d}{dt_1} \big(e^{-(t_0-t_1) (\mathrm{H}_0-V)} 
e^{-t_1 \mathrm{H}_0}\big) \nonumber 
\\
&=
 e^{-t_0 \mathrm{H}_0} + \int_0^{t_0}  \!\!dt_1 \; 
\big(e^{-(t_0-t_1) (\mathrm{H}_0-V)} 
V e^{-t_1 \mathrm{H}_0}\big) \nonumber 
\\
&= e^{-t_0 \mathrm{H}_0} + \int_0^{t_0} \!\! dt_1 \; 
\big(e^{-(t_0-t_1) \mathrm{H}_0} 
V e^{-t_1 \mathrm{H}_0}\big) \nonumber
\\
& + \int_0^{t_0} \!\! dt_1 \int_0^{t_0-t_1} \!\! dt_2 
\; \big(e^{-(t_0-t_1-t_2) \mathrm{H}_0} 
V e^{-t_2 \mathrm{H}_0}V e^{-t_1 \mathrm{H}_0}
\big) +\dots \nonumber 
\\
& + \int_0^{t_0} \!\!\! dt_1 \dots \!
\int_0^{t_0-t_1-\dots-t_{n-1}} \!\!\!\!\!\! dt_n 
\; \big(e^{-(t_0-t_1-\dots-t_n) \mathrm{H}_0} 
(V e^{-t_n \mathrm{H}_0})\cdots (V e^{-t_1 \mathrm{H}_0})
\big) 
+\dots \nonumber
\\
&= e^{-t_0 \mathrm{H}_0} + \sum_{n=1}^\infty t_0^n \int_{\Delta^n} d^n\alpha
\Big(e^{-t_0(1-|\alpha|)\mathrm{H}_0}
\prod_{j=1}^n (V e^{-t_0\alpha_j \mathrm{H}_0})\Big) \;,
\nonumber
\end{align}
where the integration is performed over the standard $n$-simplex
$\Delta^n :=\{\alpha:=(\alpha_1,\dots,\alpha_n) \in \mathbb{R}^n\;,~
\alpha_i \geq 0\;,~ |\alpha|:=\alpha_1+\dots+\alpha_n \leq 1\}$.

Using $\mathrm{tr}(e^{\omega\Sigma t})=(2 \cosh (\omega t))^4$ and the
Mehler kernel (\ref{Mehler-n}), 
the vacuum contribution without $V$ is 
\begin{align}
\mathrm{Tr}(e^{-t(H+\omega\Sigma)\otimes 1_2}) &=
\big(2\;\mathrm{tr}(e^{\omega\Sigma t}) \big)
\int_{\mathbb{R}^4} dx \; e^{-tH}(x,x)
\\ 
&= 2(2\cosh(\omega t))^4 \cdot
\Big(\frac{\omega}{2\pi\sinh(2\omega t)}\Big)^2 
\int_{\mathbb{R}^4} dx  \;e^{-\omega \tanh(\omega t) \|x\|^2}
\nonumber
\\
&=\frac{2}{\tanh^4(\omega t)}\;.
\nonumber
\end{align}
With $\coth^4(\omega t)= \frac{1}{(\omega t)^4} + \frac{4}{3 (\omega
  t)^2} + \frac{26}{45} + \mathcal{O}(t^2)$ we get under the usual
assumption $\chi^{(k)}(0)=0$ for $k=1,2,3,\dots$ the asymptotic
expansion\footnote{The Laplace transformation for the vacuum
  contribution can be performed exactly. For powers
  of $\coth x= \frac{1+e^{-2x}}{1-e^{-2x}}$ we have 
\begin{align*}
\Big(\frac{1+y}{1-y}\Big)^n 
&= 1+  \sum_{k=1}^\infty  \underbrace{\frac{(k+n-1)!}{k!} \,{}_2F_1\Big( 
\genfrac{}{}{0pt}{}{-k\,,\;-n}{1-k-n}\Big|-1\Big)}_{= F_n(k)} y^k\;.
\end{align*}
Particular values are $F_1(k)=2$, $F_2(k)=4k$, 
$F_3(k)=8k^2+4$, $F_4(k)=16k^3+32k$ and 
$F_5(k)=32k^4+160k^2+48$.
Inserted into (\ref{SpecAc}) we obtain after Laplace transformation
\begin{align*}
S_0(\mathcal{D}_A) =  2\chi(0)
+\sum_{k=1}^\infty (32 k^3+64k) \chi(2\omega k)\;.
\end{align*}}
\begin{align}
S_0(\mathcal{D}_A)  &= 
\frac{2 \chi_{-4} }{\omega^4} + \frac{8 \chi_{-2} }{3\omega^2} + 
\frac{52 \chi_{0} }{45} \;. 
\end{align}

For the further computation we distinguish the vertices (see
(\ref{DA2}), (\ref{Dphi}) and (\ref{FA}))
\begin{align}
\label{vertices}
V_1 &:= \mathrm{diag}\big(\mathrm{i} \{\partial^\mu ,A_\mu \} \otimes 1,
\mathrm{i} \{\partial^\mu ,B_\mu \} \otimes 1\big)\;, 
\\
V_2 &:= \mathrm{diag}\big(- A_\mu A^\mu \otimes 1 - |\phi^2|\otimes 1,
- B_\mu B^\mu \otimes 1 - |\phi^2|\otimes 1\big)\;, 
\nonumber
\\
V_3 &:= \mathrm{diag}\big(
- \mathrm{i}F^A_{\mu\nu}  \otimes \tfrac{1}{4} [b^{\dag\mu}
-b^\mu,b^{\dag\nu} -b^\nu] ,\ 
- \mathrm{i}F^B_{\mu\nu}  \otimes \tfrac{1}{4} [b^{\dag\mu}
-b^\mu,b^{\dag\nu} -b^\nu] \big)\;, 
\nonumber
\\
V_4&= \left( \begin{array}{cc} 0 & -D_\mu \phi \otimes (b^{\dag\mu}
    -b^\mu) \\
\overline{D_\mu \phi} \otimes (b^{\dag\mu}
    -b^\mu) & 0 \end{array} \right)\;.
\nonumber
\end{align}

We compute the traces of the spectral action in the same way as
the residues of the $\zeta$-function in Appendix~\ref{appendix}. 
The main step consists in computing the following trace:
\begin{align}
S_{t_1,\dots,t_v}(\tilde{V}_1,\dots,\tilde{V}_v) :=
\mathrm{Tr}\Big( \tilde{V}_1 e^{-t_1 H} \tilde{V}_2 e^{-t_2 H} \dots 
\tilde{V}_v e^{-t_v H}\Big)\;,
\end{align}
either with $\tilde{V}_i=f_i$ or $\tilde{V}_i
=-\mathrm{i} \{\partial_\mu,f_i^\mu\}
=-\mathrm{i}(\partial_\mu f^\mu_i)-2\mathrm{i} f_i^\mu \partial_\mu$. We
realise this alternative as $\tilde{V}_i=f_i^{1-n_i}\{-\mathrm{i}
\partial_\mu,f^\mu\}^{n_i}$ with $n_i \in \{0,1\}$:
\begin{align}
&S^{n_1\dots n_v}_{t_1,\dots t_v}(f_1,\dots,f_v) 
\\*
&= \sum_{k_1=0}^{n_1}\dots\sum_{k_v=0}^{n_v} \omega^{k_1+\dots+k_v}
\int_{(\mathbb{R}^4\times \mathbb{R}^4)^v} 
\Big(\prod_{i=1}^v\frac{dx_i dp_i}{(2\pi)^4} \Big)
\nonumber
\\*
& \times \Big(\prod_{i=1}^v \hat{f}_i^{1-n_i}(p_i) 
\Big(\hat{f}_i^{\mu_1}(p_i) 
p_{i,\mu_i}^{1-k_i} P^{k_i}_{\mu_i}\Big(2\omega t_i, \frac{\partial}{\partial p_i},
\frac{\partial}{\partial p_{i+1}}\Big) \Big)^{n_i} \Big)
\Big(\prod_{i=1}^v e^{-t_i H}(x_i,x_{i+1}) e^{\mathrm{i} p_i x_i}\Big)
\nonumber
\\
&=\sum_{k_1=0}^{n_1}\dots\sum_{k_v=0}^{n_v} 
\int_{(\mathbb{R}^4)^v} 
\Big(\prod_{i=1}^v\frac{dp_i}{(2\pi)^4} \Big)
\frac{\omega^{k_1+\dots+k_v}}{(2 \sinh(\omega(t_1+\dots+t_v)))^4} 
\nonumber
\\
& \times \Big(\prod_{i=1}^v \hat{f}_i^{1-n_i}(p_i) 
\Big(\hat{f}_i^{\mu_1}(p_i) 
p_{i,\mu_i}^{1-k_i} P^{k_i}_{\mu_i}\Big(2\omega t_i, \frac{\partial}{\partial p_i},
\frac{\partial}{\partial p_{i+1}}\Big) \Big)^{n_i} \Big)e^{-\frac{1}{4} pQ^{-1}p}
\;,
\nonumber
\end{align}
where $P_\mu$ and $Q^{-1}$ are given in (\ref{P}) and (\ref{invQ}).
From the formulae analogous to (\ref{contr-single}) and
(\ref{contr-different}) we thus obtain
\begin{align}
&S^{n_1\dots n_v}_{t_1,\dots t_v}(f_1,\dots,f_v) 
\label{Stf}
\\
&= \sum_{\mbox{\tiny$
\begin{array}{c}
k_1{+}r_{11}{+}\dots{+} r_{1v}=n_1,\dots,\\
k_1{+}r_{v1}{+}\dots{+} r_{vv}=n_v,\\
r_{ii}=0\;,~r_{ij}=r_{ji}
\end{array}$}} \!\!\!\!\! 
\int_{(\mathbb{R}^4)^v} 
\Big(\prod_{i=1}^v\frac{dp_i}{(2\pi)^4} \Big)
\frac{1}{(2 \sinh(\omega t))^4} 
\Big(\prod_{i=1}^v \hat{f}_i^{1-n_i}(p_i) 
\big(\hat{f}_i^{\mu_i}(p_i)\big)^{n_i} \Big)
\nonumber
\\
& \times 
\Big(\prod_{i=1}^v \Big( \sum_{j \neq i} 
\frac{\sinh (\omega t_{ji})}{\sinh (\omega t)} p_{j,\mu_i}\Big)^{k_i}\Big)
\Big( \prod_{i \leq j} \Big(2\omega \delta_{\mu_i\mu_j}
\frac{\cosh (\omega t_{ji})}{ \sinh (\omega t)} \Big)^{r_{ij}}
\Big) e^{-\frac{1}{4} pQ^{-1}p} \;,
\nonumber
\end{align}
where $t_{ji}:=t_j+\dots+t_{i-1}-t_i-\dots -t_{j-1}$ and $t:=t_1+\dots+t_v$

For the spectral action we are interested in the small-$t$ behaviour.
From (\ref{invQ}) we know that the singularity in
$\sinh^{-4-\sum_{i<j} r_{ij}}(\omega t)$ is protected by
$\exp(-\frac{(p_1+\dots + p_v)^2}{4\omega\tanh (\omega t)})$ unless the total
momentum is conserved. Thus, Taylor-expanding the prefactor about
$p_v=-(p_1+\dots + p_{v-1})$ up to order $\rho$ and Gau\ss{}ian
integration in $p_v$ yields
\[
S^{n_1,\dots n_v}_{t_1,\dots,t_v} = 
\mathcal{O}(t^{-2-\lfloor \frac{n_1+\dots+n_v}{2}\rfloor 
+ \lceil \frac{\rho}{2}\rceil})\;. 
\]
To obtain the spectral action, there are apart from the (at most)
$t$-neutral matrix trace the $v$ integrations over $t_1,\dots,t_v$
which contribute another power of $t^v$. If there are $v_i$ vertices
of type $V_i$ present, with $v_1+\dots+v_4=v$, then
$n_1+\dots+n_v=v_1$, and we have for such a contribution 
\[
S_t(V_1^{v_1} \dots V_4^{v_4}) = 
\mathcal{O}(t^{-2+v_2+v_3+v_4+\lceil \frac{v_1}{2}\rceil 
+ \lceil \frac{\rho}{2}\rceil})\;. 
\]
Only the non-positive exponents contribute to the asymptotic expansion
so that 
it suffices to compute the following traces of vertex combinations:
\begin{enumerate}
\item $V_2$ with Taylor expansion up to order $\rho=2$ 
($V_3$ and $V_4$ are traceless, and in $V_1$ alone there is necessarily 
$k_1=n_1=1$ and then no sum over $i \neq j$),

\item $V_1V_1$ with Taylor expansion up to order $\rho=2$, 

\item $V_1 V_2$, $V_2V_1$ and $V_1 V_1 V_2$,  $V_1 V_2 V_1$,  $V_2 V_1 V_1$   
with Taylor expansion up to order $\rho=0$,

\item $V_2V_2$, $V_3V_3$ and $V_4 V_4$ with Taylor expansion 
up to order $\rho=0$ (mixed products are traceless), 

\item $V_1V_1V_1$ and $V_1V_1V_1V_1$ with Taylor expansion up 
to order $\rho=0$. 

\end{enumerate}

We compute these vertex combinations in Appendix~\ref{App-B}. The
spectral action is the sum of (\ref{S2}), (\ref{S11}), 
(\ref{S22}), (\ref{S12}), (\ref{S112}), (\ref{S111}) and (\ref{S1111}).
Altogether, the spectral action of
the Abelian Higgs model reads 
\begin{align}
S(\mathcal{D}_A) &= 
  \frac{2 \chi_{-4} }{\omega^4} + \frac{8 \chi_{-2} }{3\omega^2} + 
\frac{52 \chi_{0} }{45} 
\\
&+ \frac{\chi_0}{\pi^2} 
\int_{\mathbb{R}^4} d x\;\Big\{
\frac{5}{12} (F^{\mu\nu}_A F_{A\mu\nu}+ 
F^{\mu\nu}_B F_{B\mu\nu})
+ \overline{D_\mu \phi} (D^\mu \phi)
\nonumber
\\*
& \qquad\quad\qquad- \frac{2\chi_{-1}}{\chi_0} |\phi|^2 + |\phi|^4 
+ 2\omega^2 \|x\|^2 |\phi|^2\Big\}(x)\;.
\nonumber
\end{align}
The scalar sector (putting $A=B=0$ and ignoring the constant) 
is almost identical to the commutative version of the 
renormalisable $\phi^4$-action \cite{Grosse:2004yu},
\begin{align}
S(\mathcal{D}_A)|_{A=b=0} &= 
\frac{\chi_0}{\pi^2} 
\int_{\mathbb{R}^4} d x\;\Big\{
\partial_\mu \bar{\phi} (\partial^\mu \phi)
+ 2\omega^2 \|x\|^2 |\phi|^2
- \frac{2\chi_{-1}}{\chi_0} |\phi|^2 + |\phi|^4 
\Big\}(x)\;.
\end{align}
The crucial difference is the negative mass square term, which leads
to a drastically different vacuum structure, as shown in the next
section.

\section{Field equations}

\label{sec:eq-motion}

\begin{figure}[h]
\begin{picture}(75,42)
\put(0,0){\includegraphics[width=7.5cm,bb=100 480 490 720]{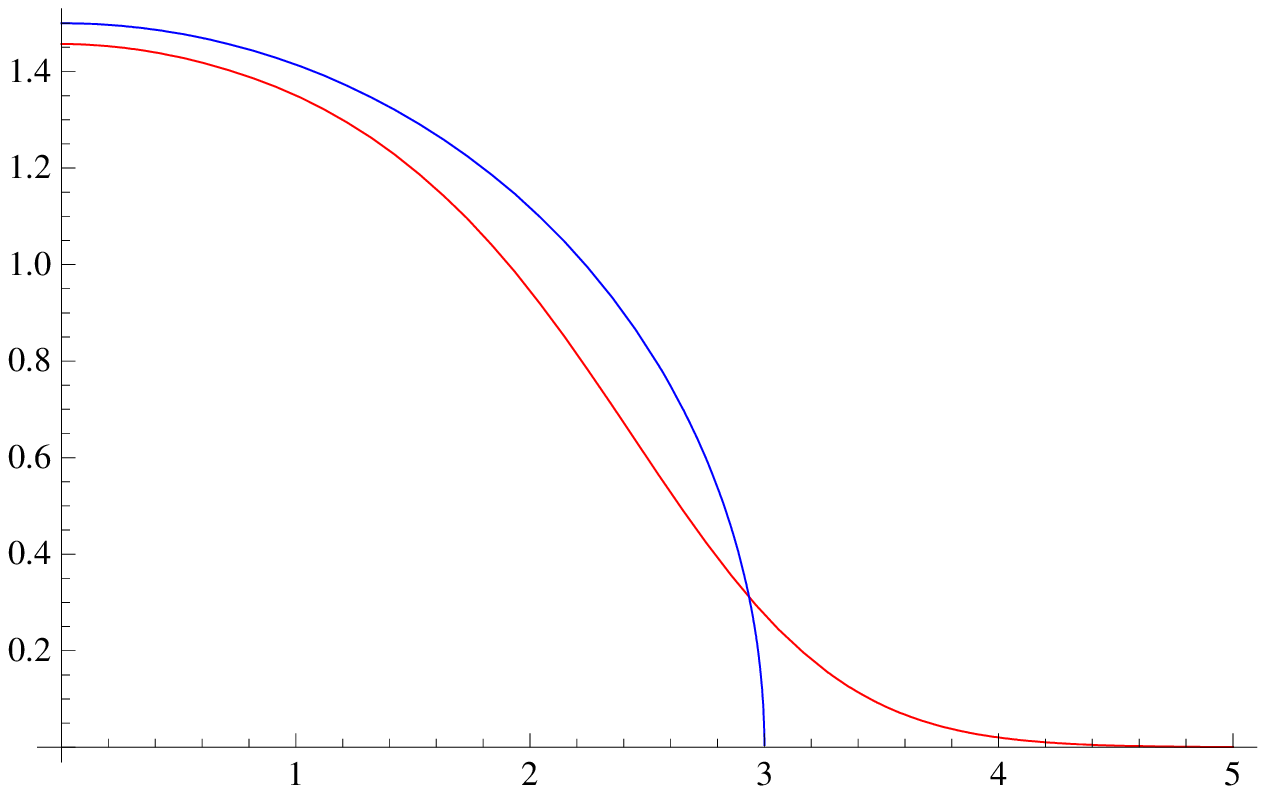}}
\put(10,10){\footnotesize$\mu=1.5$}
\end{picture}
\begin{picture}(75,42)
\put(0,0){\includegraphics[width=7.5cm,bb=100 480 490 720]{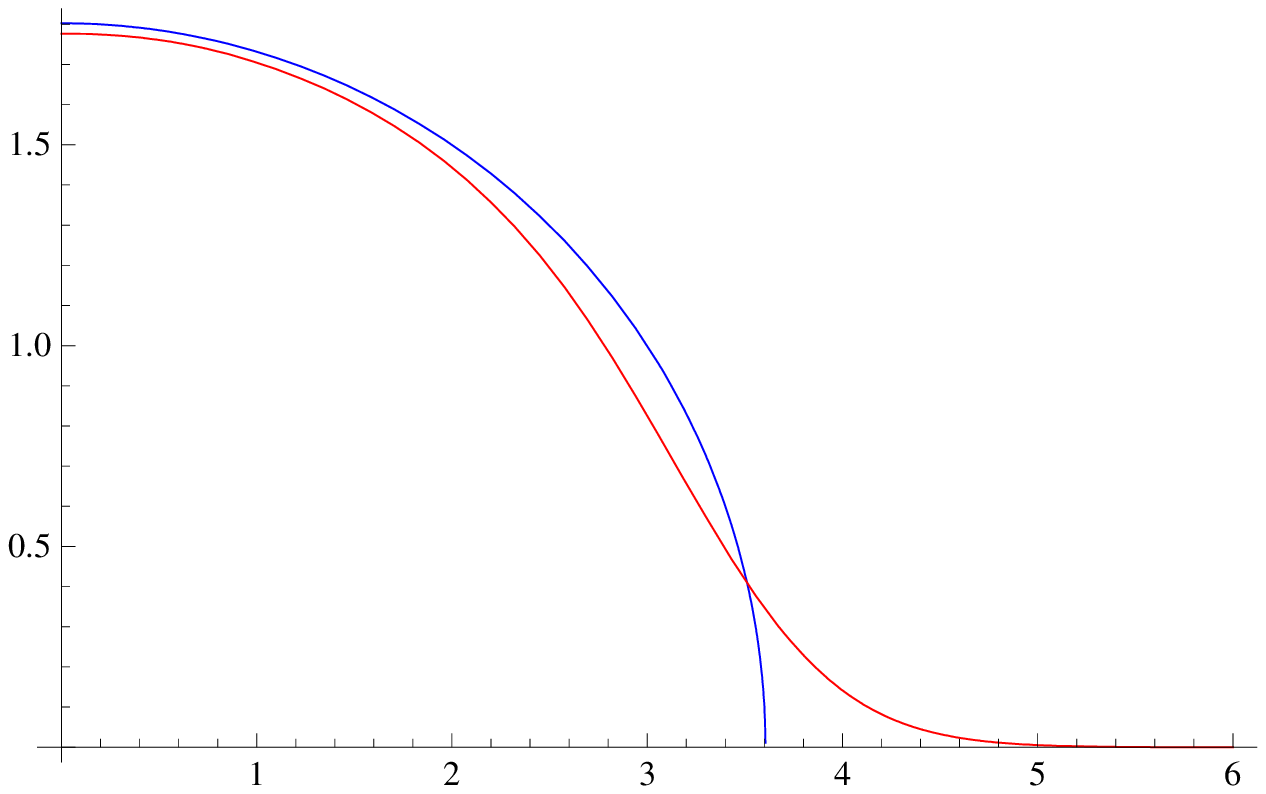}}
\put(10,10){\footnotesize$\mu=1.802\dots$}
\end{picture}
\caption{The lower curve at $r=0$ shows $\varphi_{vac}(r)$ in units of
  $\frac{2}{\sqrt{\lambda}}$, with cut-off at $N=10$. The upper curve
  at $r=0$ is the ellipse $\varphi^2+\frac{1}{4}r^2=\mu^2$. The error is below
  $1\%$ for $r<1.8\mu$. The true curve $\varphi_{vac}(r)$
is expected to stay always below the ellipse and to connect smoothly
(at least $C^2$) to $\varphi_{vac}=0$ for $r>2\mu$. 
\label{fig1}}
\end{figure}

We can assume the solution of the corresponding equation of motion to
be given by $A=B=0$ and $\phi$ a real function. Then, the Euler-Lagrange
equation reads
\begin{align}
-\Delta \phi + 2\omega^2 \|x\|^2 \phi  +2 \phi^3 - 2
\tfrac{\chi_{-1}}{\chi_0} \phi=0\;.
\end{align}
In terms of the rescaled radius $r=2^{\frac{1}{4}}\sqrt{\omega} 
\|x\|$ and the rescaled field $\phi=\frac{\pi}{\sqrt{2} \chi_0}
\varphi$ 
we have the  rotationally invariant equation
\begin{align}
&-\varphi''(r)-\frac{3}{r} \varphi'(r) + (r^2-4\mu^2) \varphi(r) = - \lambda
\varphi^3(r)\;,
\\
&\mu^2 = \frac{\chi_{-1}}{\sqrt{8} \omega \chi_0}\;,\quad
\lambda = \frac{\pi^2}{\sqrt{2}\omega \chi_0}\;.
\nonumber
\end{align}
We expand $\varphi$ in terms of eigenfunctions of the four-dimensional
harmonic oscillator, 
\begin{align}
\varphi &= \frac{2}{\sqrt{\lambda}}
\sum_{n=0}^\infty c_n \varphi_n\;, 
\\ 
\nonumber
\varphi_n &:= e^{-\frac{r^2}{2}} L_n^1(r^2)\;,\qquad 
\Big( -\frac{d^2}{dr^2}-\frac{3}{r} \frac{d}{dr} + r^2 \Big)\varphi_n= 
4(n+1) \varphi_n \;.
\end{align}
We are thus left with the equation
\begin{align}
\sum_{n=0}^\infty c_n(\mu^2- n-1) \varphi_n 
= \sum_{k,l,m=0}^\infty c_k c_l c_m 
\varphi_k \varphi_l \varphi_m 
\end{align}
or, using the orthogonality relation,
\begin{align}
c_n(\mu^2{-}n{-}1) = \frac{1}{(n+1)} \sum_{k,l,m=0}^\infty \!\!
  c_k c_l c_m 
\int_0^\infty d t\;e^{-2t} \,t\,L^1_k(t)\,
L^1_l(t)\,
L^1_m(t)\,
L^1_n(t)\;.
\end{align}
The generating function $\displaystyle (1-z)^{-\alpha-1}\exp(-\frac{xz}{1-z})
=\sum_{k=0}^\infty L^\alpha_k(t) z^k$ is used to obtain
\begin{align}
&c_n(\mu^2-n-1) 
\\
& = \sum_{k,l,m=0}^\infty 
  \frac{c_k c_l c_m }{k!l!m!} \Big(\frac{d^k}{dw^k}
\frac{d^l}{dy^l}\frac{d^m}{dz^m}
\frac{(1-yz-yw-wz+2wyz)^n}{(2-y-z-w+yzw)^{n+2}}
\Big)_{w=y=z=0}\;.
\nonumber
\end{align}
With a cut-off $N$ for the matrix indices, this equation can be solved
numerically. It turns out that except for a region about $r=4\mu^2$
the convergence is quite good. Figure~\ref{fig1} contains plots of the
vacuum solution $\varphi_{vac}(r)$ for $4\mu^2=9$ and $4\mu^2=13$
compared with the ellipse $\varphi^2+\frac{1}{4}r^2=\mu^2$.
We learn that $\varphi_{vac}(r) < \sqrt{\mu^2-\frac{1}{4} r^2}$ due to
the negative curvature
$\frac{1}{\varphi}(\varphi''+\frac{3}{r}\varphi')<0$ which effectively
reduces $\mu^2$. For $r>2\mu$ we should have $\varphi_{vac}(r)=0$ as
the only solution\footnote{The numerical convergence in the figure is
  bad for $r\approx 2\mu$.}.  We also expect that for $\mu \to \infty$,
where the ellipse becomes flat, the vacuuum solution approaches its
limiting ellipse.  This limit is connected to the limit $\omega \to
0$, i.e\ $r=2^{\frac{1}{4}} \sqrt{\omega} \|x\|\to 0$. In this the
limit the usual constant Higgs vacuum is recovered:
\begin{align}
\lim_{\omega \to 0} \phi^2
=\frac{\pi^2}{2 \chi_0^2} 
\frac{4 \mu^2}{\lambda} = \frac{\chi_{-1}}{\chi_0^2}=\text{const}\;.
\end{align}
For finite $\omega$ the cut-off for $\varphi_{vac}$ at $r=2\mu$ implies
that $\varphi_{vac}$ is an integrable function. 

\begin{figure}[h]
\begin{picture}(75,43)
\put(0,0){\includegraphics[width=7.5cm,bb=100 480 490 720]{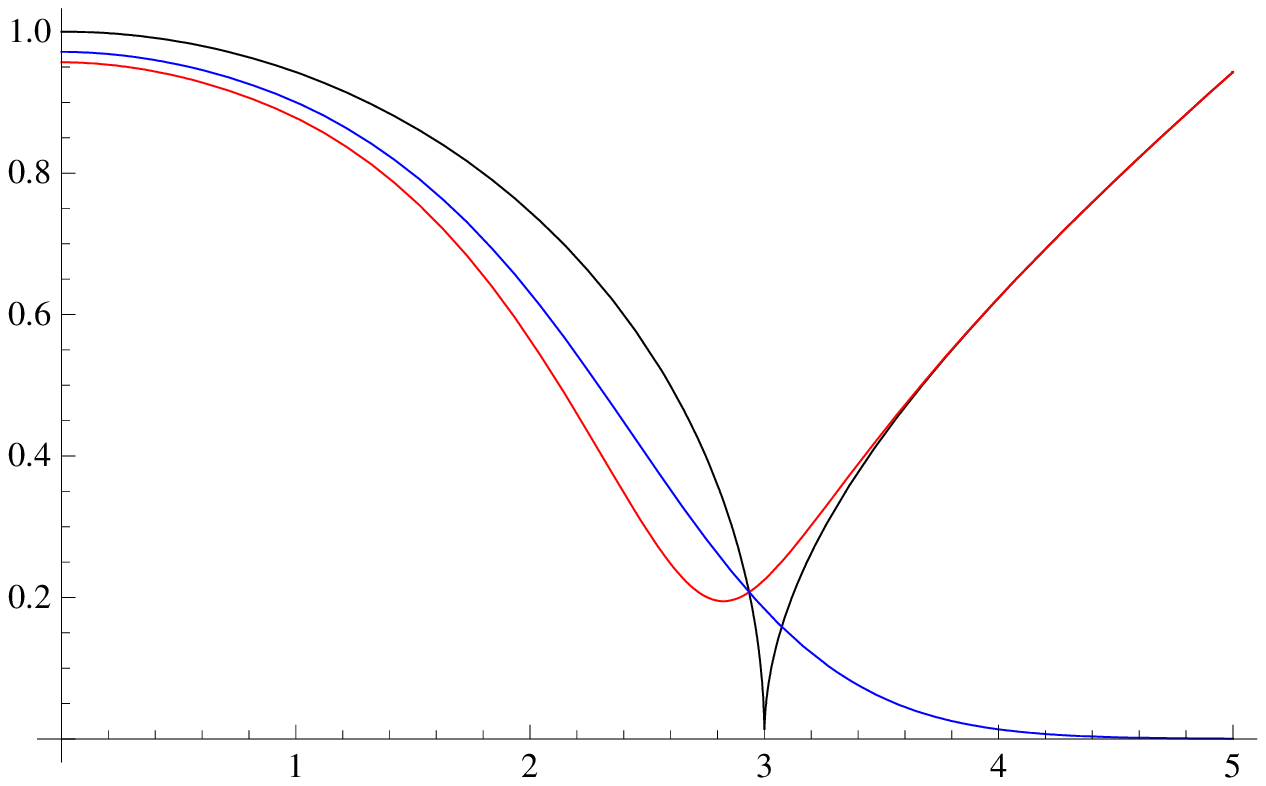}}
\put(10,17){\scriptsize$\sqrt{1{-}\frac{r^2}{4\mu^2}}$}
\put(56,17){\scriptsize$\sqrt{\frac{r^2}{8\mu^2}{-}\frac{1}{2}}$}
\put(10,7){\footnotesize$\mu=1.5$}
\end{picture}
\begin{picture}(75,43)
\put(0,0){\includegraphics[width=7.5cm,bb=100 480 490 720]{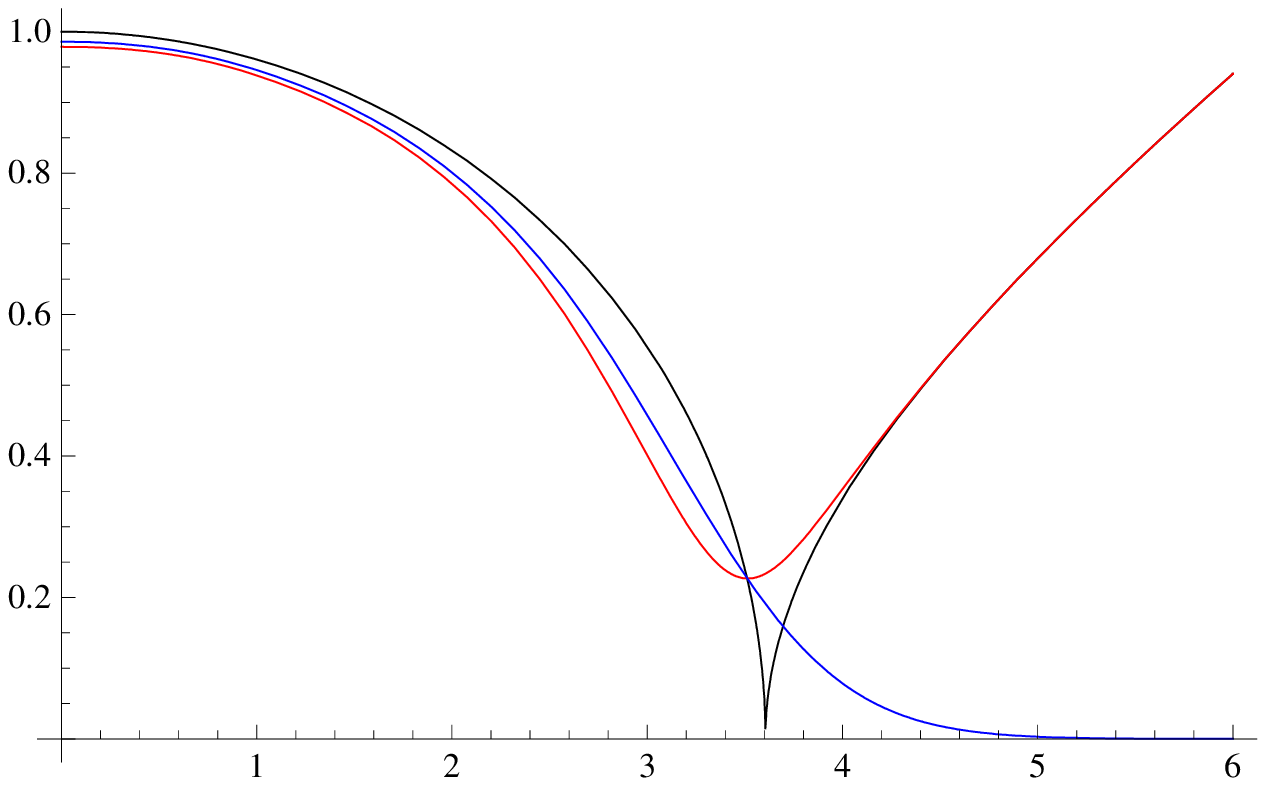}}
\put(10,17){\scriptsize$\sqrt{1{-}\frac{r^2}{4\mu^2}}$}
\put(56,17){\scriptsize$\sqrt{\frac{r^2}{8\mu^2}{-}\frac{1}{2}}$}
\put(10,7){\footnotesize$\mu=1.802\dots$}
\end{picture}
\caption{The scale $\frac{\varphi_{vac}}{\mu}(r)$ (middle curve at $r=0$)
 of the gauge field mass compared with
the scale $\frac{1}{\mu}\sqrt{\frac{3}{2}\varphi_{vac}^2(r)-\frac{1}{2}\mu^2
+\frac{1}{8}r^2}$ of the Higgs field mass (lowest curve at $r=0$) and
the limiting ellipse $s^2+\frac{r^2}{4\mu^2}=1$ and hyperbola 
$\frac{r^2}{8\mu^2}-s^2=\frac{1}{2}$. Cut-off again at $N=10$. The
true curve  $\frac{\varphi_{vac}}{\mu}(r)$ should always stay below the
ellipse and connect smoothly to  $\frac{\varphi_{vac}}{\mu}=0$ for
$r>2\mu$. The true curve  
$\frac{1}{\mu}\sqrt{\frac{3}{2}\varphi_{vac}^2(r)-\frac{1}{2}\mu^2
+\frac{1}{8}r^2}$ should stay below $\frac{\varphi_{vac}}{\mu}$ for
$r<2\mu$, whereas for $r>2\mu$ one should exactly have 
$\frac{1}{\mu}\sqrt{\frac{3}{2}\varphi_{vac}^2-\frac{1}{2}\mu^2
+\frac{1}{8}r^2}(r)=\sqrt{\frac{r^2}{8\mu^2}{-}\frac{1}{2}}$.
\label{fig2}}
\end{figure}

The vacuum solution 
\begin{align}
\frac{2}{\sqrt{\lambda}}\varphi_{vac}=\sqrt{\frac{4\mu^2}{\lambda}}
\frac{\varphi_{vac}}{\mu}=\sqrt{\frac{2\chi_{-1}}{\pi^2}}
\frac{\varphi_{vac}}{\mu}
\end{align}
sets the scale for the bare masses of gauge fields and fermions. On the
other hand, the bare mass of the Higgs field is obtained from the
shift of the Higgs potential into its minimum and therefore reads 
\begin{align}
\sqrt{\sqrt{2}\omega ((r^2-4\mu^2)
+12 \mu^2 \frac{\varphi_{vac}^2}{\mu^2})
}=\sqrt{\frac{4\chi_{-1}}{\chi_0}}
\frac{\sqrt{\frac{3}{2}\varphi_{vac}^2-\frac{1}{2}\mu^2
+\frac{1}{8}r^2}}{\mu}\;.
\end{align}
We compare in Figure~\ref{fig2} 
the scale $\frac{\varphi_{vac}}{\mu}$ of gauge field mass with
the scale $\frac{1}{\mu}\sqrt{\frac{3}{2}\varphi_{vac}^2-\frac{1}{2}\mu^2
+\frac{1}{8}r^2}$ of the bare Higgs mass.
Reinserting $\omega$ we obtain 
the following \emph{two-phase structure}: 
\begin{itemize}
\item \emph{A spontaneously broken phase for $\omega^2\|x\|^2 <
    \frac{\chi_{-1}}{\chi_0}$.}
\\
Fermions, gauge fields and Higgs field are all massive, with the Higgs
mass slightly smaller than the prediction from noncommutative geometry
\cite{Chamseddine:2006ep}. In particular, this phase is the
 only existing one in the limit $\omega \to 0$, and in this limit the
 NCG prediction is recovered. 

\item  \emph{An unbroken phase for $\omega^2\|x\|^2 >
    \frac{\chi_{-1}}{\chi_0}$.}
\\
Fermions and gauge fields are massless, whereas the Higgs field
remains massive. 
\end{itemize}

The model we have studied is a toy model. But as it is a noncommutative
geometry like that of the NCG-formulation of the Standard Model
\cite{Chamseddine:2006ep}, it is ultimately an experimental question
to set limits on the frequency parameter $\omega$. To be compatible
with both high energy and cosmological data, $\omega$ has to be
extremely small. We definitely live in the spontaneously broken phase
$\omega^2\|x\|^2 < \frac{\chi_{-1}}{\chi_0}$, and the observable
universe is very close to $\omega^2\|x\|^2=0$. Nevertheless, a
regulating $\omega \neq 0$ has some nice consequences such as 
integrability of the Higgs vacuum and integrability of the
cosmological constant. 

One may speculate how an $\omega \neq 0$ can be detected. We mentioned
the reduction of the ratio between Higgs mass and $Z$ mass compared
with the NCG prediction. However, in presence of $\omega \neq 0$ the
$\beta$-functions must be recomputed so that at the moment no
prediction is possible. In cosmology, limits for $\omega$ could be
obtained from precision measurements of the ratio between the proton
mass and the electron mass at far distance.  The electron mass which
governs the atomic spectra via the Rydberg frequency should vary in
the same way as the Higgs scale $\frac{\varphi_{vac}}{\mu}$. On the
other hand, the proton mass arises mainly from broken scale invariance
in QCD and therefore can be regarded as constant. This means that the
gravitational energy of a standard star is constant whereas its
transition into radiation energy might vary with the position of the
star in the universe. Observational limits on such a variation would
limit the value of $\omega$.

Another observable consequence could be a variation of the cosmological
constant. The Higgs potential at the vacuum solution is negative and
hence reduces the volume term of the cosmological constant. Thus, the
effective cosmological constant would increase with the radius (the
masses of gauge fields and fermions 
dissipate into the cosmological constant).

\section{Conclusion and perspectives}

We have proposed a definition for non-compact spectral triples
$(\mathcal{A},\mathcal{H},\mathcal{D})$ where the algebra is
allowed to be non-unital but the resolvent of the operator
$\mathcal{D}$ remains compact. The metric dimension is defined via the
dimension spectrum; it is (in general) different from the noncommutative 
dimension given by the decay rate of the characteristic values of the
resolvent. 

Our definition excludes non-compact manifolds with the standard Dirac
operator, but this is necessary for a well-defined index problem and a
well-defined spectral action in the non-compact case. An example for
our definition is given by operators $\mathcal{D}$ which are 
square roots of the $d$-dimensional harmonic oscillator Hamiltonian
$-\Delta +\omega^2 x^2$. These square roots are constructed by
conjugation of the partial derivatives with $e^{\pm \omega h}$, where
$h$ is the Morse function. This relates to supersymmetric quantum
mechanics, in particular to a special case of Witten's work
\cite{Witten:1982im} on Morse theory.

The most involved piece of work was the computation of the dimension
spectrum which showed that the metric dimension is the oscillator
dimension and that all residues of the operator zeta function are
local. Due to its relation to supersymmetry, there are in fact two
Dirac operators $\mathcal{D}_1$ and $\mathcal{D}_2$, which define two
distinct images $\boldsymbol{\gamma}_1$ and $\boldsymbol{\gamma}_2$ of
the $d$-dimensional volume form, and only the product
$\boldsymbol{\gamma}_1\boldsymbol{\gamma}_2$ defines the
$\mathbb{Z}_2$-grading.

We have computed the spectral action for the corresponding Connes-Lott
two-point model. In distinction to standard $\mathbb{R}^d$, the
spectral action is finite also in the cosmological constant part. The
result is an Abelian Higgs model with additional harmonic oscillator
potential for the Higgs field. The resulting field equations show a
phase transition phenomenon: There is a spontaneously broken phase
below a critical radius determined by the oscillator frequency
$\omega$, which for small enough $\omega$ is qualitatively identical
to standard Higgs models. Possible observable consequences are
discussed at the end of the previous section. Above the critical
radius we have an unbroken phase with massless gauge fields. This
phase is necessary to have an integrable vacuum solution for the Higgs
field.

The class of spectral triples we proposed deserves further
investigation. We show with H.~Grosse \cite{Grosse:2009??} that there
is an isospectral Moyal deformation of the harmonic oscillator
spectral triple. Some ideas appeared already in our preprint
\cite{Grosse:2007jy}, but the mathematical structure was unclear at
that point.  The field equations of the preprint \cite{Grosse:2007jy}
are correct, but their ``solution'' is wrong. It misses the phase
transitions which we first observed for the commutative model in the
present paper. We expect that the phase structure is much richer in
the Moyal-deformed model. A hint can already be found in the pure gauge
field sector, which leads in terms of ``covariant coordinates'' to the
field equation $[X^\mu ,[X_\mu,X_\nu]]=0$. This equation has the Moyal
deformation $[X_\mu,X_\nu]=\mathrm{i}\Theta_{\mu\nu}=\text{const}$ as
a solution, but also commutative coordinates $[X_\mu,X_\nu]=0$; the
preferred solution arises from a subtle interplay with the boundary
conditions. One may speculate that these boundary conditions change
with the temperature of the universe, so that the (non)commutative
geometry could emerge through a cascade of phase transition when the
universe cools down. The Moyal-deformed harmonic oscillator spectral
triple could serve as an excellent toy model to study these
transitions.

On the mathematical side, the relation to supersymmetric quantum
mechanics needs further study. In particular, a real structure (or
better several real structures) must be identified to reduce the
multiplicity of the action of the algebra from its present value $2^d$
to $2^{\frac{d}{2}}$ in order to support a spin$^c$ structure. One
should also allow for a non-trivial projection $e$ to define 
the smooth subspace
$\mathcal{H}_\infty =e\mathcal{A}^n$ of the Hilbert space. The
corresponding action of $\mathcal{D}_i$ or its components
$\mathfrak{Q},\mathfrak{Q}^\dag$ would then permit a complete
reformulation of Witten's approach \cite{Witten:1982im} to Morse
theory in the framework of spectral triples and noncommutative index
theory.

\section*{Acknowledgements}

I would like to thank the Clay Mathematics Institute for inviting me
to write this contribution to the birthday volume for Alain Connes.

It is a pleasure to thank Harald Grosse for the long-term
collaboration in which we found the need for $\omega$ in
noncommutative scalar models and the possibility of a spectral triple
for these models. The relation to supersymmetric quantum mechanics and
Morse theory appeared in discussions with Christian Voigt. Several
hints by Alain Connes on the $\mathbb{Z}_2$-grading were important to
identifying the structure observed in
section~\ref{sec:Orientability}. The index formula was inspired by
lectures of Toshikazu Natsume at the Tehran conference on
noncommutative geometry.

The scientific exchange was supported by the grants SFB 478
of the Deutsche Forschungsgemeinschaft and MRTN-CT-2006-031962 of
the European Union.

\begin{appendix}

\section{Proof of Theorem \ref{thm-spectrum}}
\label{appendix}

Let $\mathcal{D}$ denote $\mathcal{D}_1$ or $\mathcal{D}_2$. 
The spectral identity 
$\displaystyle 
A = \frac{1}{\pi} \int_0^\infty \frac{d\lambda}{\sqrt{\lambda}}
\frac{A^2}{A^2+\lambda}$
for a positive selfadjoint operator $A$ leads to
\begin{align}
\delta T= 
\frac{1}{\pi} \int_0^\infty d\lambda\,\sqrt{\lambda} 
\frac{1}{\langle \mathcal{D}\rangle^2 + \lambda} 
[ \mathcal{D}^2,T]  
\frac{1}{\langle \mathcal{D}\rangle^2 + \lambda}\;.
\end{align}
From (\ref{H-x}) we recall that $\mathcal{D}^2= H+\omega\Sigma$, where
$H=-\partial^\mu\partial_\mu +\omega^2 x_\mu x^\mu$ and
$\Sigma= [b^\dag_\mu,b^\mu]$ satisfy $[H,\Sigma]=0$. This
implies 
\begin{align}
\delta^n T &=
\sum_{k=0}^m \binom{n}{k} 
\big(\omega \,\mathrm{ad}(\Sigma)\big)^{n-k}\bigg(
\frac{1}{\pi^n} \int_0^\infty \prod_{i=1}^n
\frac{d\lambda_i\,\sqrt{\lambda_i}}{\langle \mathcal{D}\rangle^2 
+ \lambda_i} 
(\mathrm{ad}(H))^k (T)
\prod_{j=1}^n \frac{1}{\langle \mathcal{D}\rangle^2 {+} \lambda_j}
\bigg).
\label{deltaT}
\end{align}
The case $T=[\mathcal{D}_1,f]= \partial_\mu f \otimes (b^{\dag\mu}
-b^\mu)$ or
$T=[\mathcal{D}_2,f]= \partial_\mu f \otimes (\mathrm{i}b^{\dag\mu}
+\mathrm{i} b^\mu)$ is
also reduced to $T=f$; only $\mathrm{ad}(\Sigma)$
distinguishes them, and each application of $\mathrm{ad}(\Sigma)$ makes
$\delta T$ more regular. It is therefore sufficient to study
$T=f$ and $k=n$. Using $[H,f]=-(\Delta f)-2(\partial^\mu
f)\partial_\mu = -\{\partial_\mu,\partial^\mu f\}$, we
have
\begin{align}
\delta^n f = \sum_{k=0}^n \binom{n}{k} 2^k 
\frac{(-1)^n}{\pi^n} &\int_0^\infty \prod_{i=1}^n
\frac{d\lambda_i\,\sqrt{\lambda_i}}{\langle \mathcal{D}\rangle^2 
+ \lambda_i} 
\label{deltaT1}
\\
&\times (\Delta^{n-k}\partial^{\mu_1}\dots \partial^{\mu_k} f)
\partial_{\mu_1}\dots \partial_{\mu_k} 
\prod_{j=1}^n \frac{1}{\langle \mathcal{D}\rangle^2 + \lambda_j}\;.
\nonumber
\end{align}

By linearity, it suffices to consider $\phi=
(\delta^{n_1} f_1) \cdots (\delta^{n_v} f_v)$. 
The most convenient way is to compute $\zeta_\phi(z)$ as a trace 
over position space kernels,
\begin{align}
\label{TrbDz}
&\zeta_\phi(z) 
\\*
&:=\mathrm {Tr}\big((\delta^{n_1} f_1) \cdots 
(\delta^{n_v} f_v) \langle \mathcal{D}\rangle^{-z}\big)
 \nonumber
\\*
&=\mathrm{tr}\bigg(
\int_0^\infty \!\!\! 
dt_0 \; \frac{t_0^{\frac{z}{2}-1}}{\Gamma(\frac{z}{2})} 
 \int_{(\mathbb{R}^d)^v} \!\!\Big(\prod_{i=1}^v d y_i \Big)
(\delta^{n_1} f_1)(y_1,y_2) \cdots 
(\delta^{n_{v-1}} f_{v-1})(y_{v-1},y_v) 
\nonumber
\\*[-1ex]
& \qquad\qquad\qquad\qquad\qquad
\times (\delta^{n_{v}} f_{v})(y_{v},y_0) 
(e^{-t_0 \langle\mathcal{D}\rangle^2})(y_0,y_1)\bigg)\;.
\nonumber
\end{align}
The remaining trace $\mathrm{tr}$ is taken in  $\bigwedge(\mathbb{C}^d)$.
Further evaluation is possible thanks to the 
$d$-dimensional Mehler kernel 
 \begin{align}
e^{-t H}(x,y) 
&= \Big(\frac{\omega}{2\pi\sinh(2\omega t)}\Big)^{\frac{d}{2}}
e^{-\frac{\omega}{4} \coth(\omega t) \|x-y\|^2 
-\frac{\omega}{4}\tanh(\omega t) \|x+y\|^2 }\;,
\label{Mehler-n}
\end{align}
for $x,y \in \mathbb{R}^d$,
which solves the differential equation 
$(\frac{d}{dt} + H_{x})e^{-tH}(x,y)=0$ with initial condition
$\lim_{t \to 0} e^{-t H}(x,y)=\delta(x-y)$. Uniqueness of the
solution implies 
\begin{align}
&\int_{\mathbb{R}^d} dy \; e^{-t_1 H}(x,y) e^{-t_2 H}(y,z) 
= e^{-(t_1+t_2) H}(x,z) \;.
\label{Mehler-recombination}
\end{align}
We can therefore recombine left and right Mehler kernels 
\begin{align}
\frac{1}{\langle \mathcal{D}\rangle^2+ \lambda_{i,j_i}} 
=\int_0^\infty dt_{i,j_i}\;e^{-t_{i,j_i}(H+\omega\Sigma+1+\lambda_{i,j_i})}
  \end{align}
in 
(\ref{deltaT1}) and integrate over $\lambda_{i,j_i}$: 
\begin{align}
&(\delta^{n_i}f_i)(y_i,y_{i+1})
\\*
&= \sum_{k_i=0}^{n_i} \binom{n_i}{k_i} 2^{k_i} 
\frac{(-1)^{n_i}}{(2\sqrt{\pi})^{n_i}} 
\int_0^\infty  \prod_{j_i=1}^{n_i} \frac{d t_{i,j_i}
  ds_{i,j_i}}{(t_{i,j_i}+s_{i,j_i})^{\frac{3}{2}}} 
e^{-(1+\omega \Sigma)(S_i+T_i)} 
\nonumber
\\*
&  \times
\int_{\mathbb{R}^d} \!\!\! dx_i \, e^{-S_i H}(y_i,x_i) 
(\Delta^{n_i-k_i}\partial^{\mu^i_1}\dots \partial^{\mu^i_{k_i}} f_i)(x_i)
\frac{\partial^{k_1}}{\partial x_i^{\mu_1^i} \dots \partial x_i^{\mu^i_{k_i}}}
e^{-T_i H}(x_i,y_{i+1}) \;,
\nonumber
\end{align}
where $S_i:=\sum_{j_i=1}^{n_i} s_{j_i}$ and $T_i:=\sum_{j_i=1}^{n_i}
t_{j_i}$. We insert this into (\ref{TrbDz}), move $e^{-S_1H}$ under
the trace to the end, and perform the
$y_i$-integrations which combine the Mehler kernels into
$e^{-\frac{\tau_i}{2\omega} H}(x_i,x_{i+1})$, with 
$\tau_i=2\omega(T_i+S_{i+1}+\delta_{iv} t_0)$ and the convention 
$v+1\equiv 1$. The remaining trace in $\bigwedge (\mathbb{C}^d)$ is 
\begin{align}
\mathrm{tr}(e^{-\Sigma y})=
\mathrm{tr}(e^{-y[b^{\dag 1},b^1]} \cdots  
e^{-y[b^{\dag d},b^d]}) = (2\cosh y)^d\;.
\label{trSigma}
\end{align}
Now the $k_i$ partial derivatives of the Mehler kernel read 
\begin{align}
&\sum_{k_i=0}^{n_i} \binom{n_i}{k_i} 2^{k_i} (-1)^{n_i} 
(\Delta^{n_i-k_i}\partial^{\mu^i_1}\dots \partial^{\mu^i_{k_i}} f_i)(x_i)
\frac{\partial^{k_1}}{\partial x_i^{\mu_1^i} \dots \partial
  x_i^{\mu^i_{k_i}}}e^{-\frac{\tau_iH}{2\omega}}(x_i,x_{i+1})
\label{deltanf-2}
\\
&= \!\!\!
\sum_{k_i+2 l_i+r_i=n_i} \frac{n_i!}{l_i!k_i!r_i!} 
\omega^{n_i-k_i-l_i} (-1)^{k_i+l_i} 
2^{l_i}\coth^{l_i}(\tau_i) 
\big(\Delta^{k_i+l_i} \partial^{\mu^i_1}\dots \partial^{\mu^i_{r_i}} f_i\big)(x_i)
\nonumber
\\
&\qquad\times 
\bigg(\prod_{j=1}^{r_i} 
\big(
(x_i-x_{i+1})\coth \tfrac{\tau_i}{2} +
(x_i+x_{i+1})\tanh \tfrac{\tau_i}{2}\big)_{\mu^i_j}\bigg)
e^{-\frac{\tau_iH}{2\omega}}(x_i,x_{i+1}) \;.
\nonumber
\end{align}
We represent the $f_i$ by their Fourier transform
$f_i(x)=\displaystyle \int_{\mathbb{R}^d} \frac{dp_i}{(2\pi)^d}  \;
\hat{f}_i(p_i)\,e^{\mathrm{i}p_ix_i}$, write the $x_i,x_{i+1}$ in 
(\ref{deltanf-2}) as derivative with respect to $p_i,p_{i+1}$,
respectively, and
obtain after Gau\ss{}ian integration of the $x_i$
\begin{align}
\label{zeta}
\zeta_\phi(z) &= \sum_{\mbox{\tiny$
\begin{array}{c}k_1{+}2l_1{+}r_1=n_1,\dots,\\
k_v{+}2l_v{+}r_v=n_v
\end{array}$}}
\bigg(\prod_{i=1}^v \frac{n_i!}{l_i!k_i!r_i!} 
\omega^{n_i-k_i} \bigg)
\frac{1}{\Gamma(\frac{z}{2})(2\sqrt{\pi})^{n_1+\dots+n_v}} 
\\
& \times \int_0^\infty dt_0 \; t_0^{\frac{z}{2}-1}
\int_0^\infty \prod_{i=1}^v\prod_{j_i=1}^{n_i} \frac{dt_{i,j_i} 
ds_{i,j_i}}{(t_{i,j_i}+s_{i,j_i})^{\frac{3}{2}}} 
e^{-(t_0+\sum_{i=1}^n (S_i+T_i))}
\nonumber
\\
& \times  
\big(2\cosh \tfrac{\tau_1+\dots+\tau_v}{2}\big)^d 
\bigg(\prod_{i=1}^v \Big(\frac{2}{\omega} \coth\tau_i\Big)^{l_i}
\Big(\frac{\omega}{2\pi\sinh \tau_i}\Big)^{\frac{d}{2}} 
\bigg)
\nonumber
\\
&
\times \int_{(\mathbb{R}^d)^v} \Big(\prod_{i=1}^v \frac{d p_i}{(2\pi)^d}\Big)
 \bigg(\prod_{i=1}^v 
(p_i^2)^{k_i+l_i} p_i^{\mu_1^i}\cdots p_i^{\mu_{r_i}^i}
\hat{f}_i(p_i)\bigg)
\nonumber
\\
& \qquad\times
\bigg(\prod_{i=1}^v \prod_{j=1}^{r_i}
P_{\mu^i_j}
\Big(\tau_i; \frac{\partial}{\partial p_i},
\frac{\partial}{\partial p_{i+1}}\Big)\bigg)
\Big(
\frac{\sqrt{\pi}^{dv}\,e^{-\frac{1}{4}p Q^{-1}p}}{(\det Q)^{\frac{d}{2}}}
\Big)\;,
\nonumber
\end{align}
where 
\begin{align}
& P_{\mu^i_j}\Big(\tau_i; \frac{\partial}{\partial p_i},
\frac{\partial}{\partial p_{i+1}}\Big)
:= 
\coth \tfrac{\tau_i}{2} 
\Big(\frac{\partial}{\partial p_i^{\mu^i_j}}-
\frac{\partial}{\partial p_{i+1}^{\mu^i_j}}\Big)
+\tanh \tfrac{\tau_i}{2} 
\Big(\frac{\partial}{\partial p_i^{\mu^i_j}}+
\frac{\partial}{\partial p_{i+1}^{\mu^i_j}}\Big)
\label{P}
\end{align}
and
\begin{align}
Q&=  \frac{\omega}{2} \left({\arraycolsep 2pt \begin{array}{cccccc}
\frac{\sinh (\tau_{v}+\tau_1)}{\sinh \tau_v \sinh \tau_1} 
& \frac{-1}{\sinh \tau_1} & 0 & \dots & 0 & \frac{-1}{\sinh \tau_v} 
\\
\frac{-1}{\sinh \tau_1} & \frac{\sinh (\tau_{1}+\tau_2)}{\sinh \tau_1 
\sinh \tau_2}  & \frac{-1}{\sinh \tau_2} & \ddots & \ddots & 0 
\\
0 & \frac{-1}{\sinh \tau_2} & \frac{\sinh (\tau_{2}+\tau_3)}{\sinh \tau_2 
\sinh \tau_3}   & \ddots & \ddots & 0 
\\
\vdots & \ddots & \ddots & \ddots & \ddots & \vdots 
\\
0 & \ddots & \ddots &\ddots & 
\frac{\sinh (\tau_{v-2}+\tau_{v-1})}{\sinh \tau_{v-2} \sinh \tau_{v-1}} 
& \frac{-1}{\sinh \tau_{v-1}} 
\\
\frac{-1}{\sinh \tau_v} & 0 & 0 &\dots & 
\frac{-1}{\sinh \tau_{v-1}} & 
\frac{\sinh (\tau_{v-1}+\tau_v)}{\sinh \tau_{v-1} \sinh \tau_v} 
\end{array}}\right).
\end{align}

By Gau\ss{}-Jordan elimination and multiple use of the addition
theorems for $\sinh$ it is straightforward to compute the determinant
and the inverse of the symmetric matrix $Q$ (the result also holds for $v=1$):
\begin{align}
\det Q &=   \Big(\frac{\omega}{2}\Big)^v 
\frac{4\sinh^2(\tfrac{1}{2}(\tau_1+\dots+\tau_v))}{
\prod_{i=1}^v \sinh \tau_i}\;,
\label{detQ}
\\
(Q^{-1})_{ij}&= \frac{1}{\omega \tanh(\tfrac{1}{2}(\tau_1+\dots+\tau_v))}
+ \tilde{Q}_{ij}\;,
\label{invQ}
\\
\tilde{Q}_{ij}&= 
- \frac{2\displaystyle \sinh(\tfrac{1}{2}(\tau_i+\dots+\tau_{j-1})) 
\sinh(\tfrac{1}{2}(\tau_j+\dots+\tau_{i-1}))}{\omega
  \sinh(\tfrac{1}{2}(\tau_1+\dots+\tau_v))}\;,
\end{align}
where in $\tilde{Q}_{ij}$ one of the chains $\tau_i+\dots+\tau_{j-1}$ or
$\tau_j+\dots+\tau_{i-1}$ passes the index $v\equiv 0$. The determinant can
also be obtained from the fact that for $p=0$ we just have the trace over the
concatenation of Mehler kernels (\ref{Mehler-recombination}).

The action of $(P_{\mu^i_j})$ on $e^{-\frac{1}{4} pQ^{-1}p}$ is
partitioned into $k_i'$ out of $r_i$ single contractions, $l_i'$ double
contractions and $r_{ij}$ halves of mixed contraction with another index 
$j\neq i$ such that $k_i'+l_i'+\sum_{j\neq i} r_{ij}=r_i$. 
Their number is $\frac{r_i!}{2^{l_i'} l_i'! k_i'! r_{i1}!\cdots 
r_{iv}!}$ if we put $r_{ii}=0$ and $r_{ij}=r_{ji}$.
Together with the multiplying $p_i^{\mu^i_j}$, a
single contraction gives a factor 
\begin{align}
p_i^{\mu^i} P_{\mu^i} (-\tfrac{1}{4} pQ^{-1}p)
&= -\frac{p_i^2}{\omega} - \sum_{j\neq i} \frac{\sinh
  (\frac{\tau_i+\dots+\tau_{j-1}-\tau_j-\dots-\tau_{i-1}}{2}) }{\omega 
\sinh(\frac{\tau_1+\dots+\tau_v}{2})} p_i p_j\;.
\label{contr-single}
\end{align}
A double contraction with respect to the same index $i$ gives a factor 
\begin{align}
p_i^{\mu^i} p_i^{\nu^i} P_{\mu^i} P_{\nu^i} (-\tfrac{1}{4} pQ^{-1}p)
= \Big(-\frac{4 \coth \tau_i}{\omega} + \frac{2}{\omega} 
\coth (\tfrac{\tau_1+\dots+\tau_{v}}{2})\Big) p_i^2\;.
\label{contr-double}
\end{align}
A mixed contraction with respect to different indices $i \neq j$ gives a
factor 
\begin{align}
p_i^{\mu^i} p_j^{\mu^j} P_{\mu^i} P_{\mu^j} (-\tfrac{1}{4} pQ^{-1}p)
= 2 \frac{\cosh
  (\frac{\tau_j+\dots+\tau_{i-1}-\tau_i-\dots-\tau_{j-1}}{2}) }{\omega 
\sinh(\frac{\tau_1+\dots+\tau_v}{2})} p_i p_j\;.
\label{contr-different}
\end{align}
We insert these formulae into (\ref{zeta}) and notice that the sum over
$l_i,l_i'$ combines to a joint sum (with new index $l_i$) 
involving only the factor $\frac{p_i^2}{\omega}\coth
(\tfrac{1}{2}(\tau_1+\dots+\tau_{v}))$ from (\ref{contr-double}), whereas
$\coth \tau_i$ cancels. In the same way, the sum over $k_i,k_i'$ cancels the
term $-p_i^2$ from (\ref{contr-single}) so that only the sum over $j\neq i$
remains: 
\begin{align}
\label{zeta-P}
\zeta_\phi(z) 
&  = \!\!\!\!\! \sum_{\mbox{\tiny$
\begin{array}{c}k_1{+}2l_1{+}r_1=n_1,\dots,\\
k_v{+}2l_v{+}r_v=n_v\\
r_1{+}\dots{+}r_v=2m
\end{array}$}}
\sum_{\mbox{\tiny$
\begin{array}{c} r_{11}{+}\dots {+}r_{1v}=r_1, \dots,\\ 
r_{v1}{+}\dots {+}r_{vv}=r_v
\end{array}$}} \!\!\!\!\!
\bigg(\prod_{i=1}^v \frac{n_i!  }{l_i!k_i!} \bigg)
\frac{2^m \omega^{l_1+\dots+l_v+m}}{\Gamma(\frac{z}{2})
(2\sqrt{\pi})^{n_1+\dots+n_v}} 
\\
& \times \int_0^\infty dt_0 \; t_0^{\frac{z}{2}-1}
\int_0^\infty \prod_{i=1}^v\prod_{j_i=1}^{n_i} \frac{dt_{i,j_i} 
ds_{i,j_i}}{(t_{i,j_i}+s_{i,j_i})^{\frac{3}{2}}} 
\frac{e^{-(t_0+\sum_{i=1}^n (S_i+T_i))}}{
\big(\tanh \tfrac{\tau_1+\dots+\tau_v}{2}\big)^{d+l_1+\dots+l_v} }
\nonumber
\\
& \times  
\int_{(\mathbb{R}^d)^v} \!\!\! \Big(\prod_{i=1}^v \frac{d p_i}{(2\pi)^d}\Big)
\bigg(\prod_{i < j}
\frac{1}{r_{ij}!} \Big(\frac{\cosh\big(\frac{\tau_{j}+\dots+\tau_{i-1}}{2}
-\frac{\tau_{i}+\dots+\tau_{j-1}}{2}\big)}{ 
\sinh(\frac{\tau_1+\dots+\tau_v}{2})} p_{i} p_{j}\Big)^{r_{ij}}\bigg)
\nonumber
\\
& \times
\bigg(\!\! \prod_{i=1}^v 
\Big(\sum_{j\neq i} \frac{\sinh
  \big(\frac{\tau_j+\dots+\tau_{i-1}}{2}
-\frac{\tau_i+\dots+\tau_{j-1}}{2}\big) }{ 
\sinh(\frac{\tau_1+\dots+\tau_v}{2})} p_i p_j\Big)^{k_i}
(p_i^2)^{l_i} \hat{f}_i(p_i)\!\!\bigg)
e^{-\frac{1}{4}p Q^{-1}p} .
\nonumber
\end{align}

The zeta-function potentially has a singularity for 
$\tau=\tau_1+\dots+\tau_v\to 0$ of order 
$\tau^{\frac{z+k_1+\dots+k_v}{2}-d}$. 
The contribution $\frac{z}{2}$ is from $dt_0\;t_0^{\frac{z}{2}-1}$, 
the measure $\prod \frac{dt ds}{(t+s)^{\frac{3}{2}}}$ contributes
$\frac{n_1+\dots+n_v}{2}$, and $(\tanh
\frac{\tau}{2})^{-d-l_1-\dots-l_v} 
(\sinh \frac{\tau}{2})^{-r_{12}+\dots+ r_{v-1,v}}$ 
contribute $-(d+l_1+\dots+l_v+\frac{r_1+\dots+r_v}{2})$. 
However, the independence of the leading term in
$Q_{ij}^{-1}$ from $i,j$ shows that this singularity 
is protected by $e^{-\frac{(p_1+\dots + p_v)^2}{\omega 
\tanh \frac{\tau}{2}}}$ unless
the total momentum is conserved, $p_1+\dots + p_v=0$. The remaining
singularity is identified by a Taylor expansion in $p_v$ about
$\bar{p}_v:=-(p_1+\dots+p_{v-1})$ up to order $\rho$ to be determined later: 
\begin{align}
&F(p_1,\dots,p_v)
\label{Taylor}
\\
&= \sum_{|\alpha| \leq \rho} 
\frac{(p_v-\bar{p}_v)^\alpha}{\alpha!} 
\frac{\partial^{|\alpha|} F}{\partial p_v^\alpha}
(p_1,\dots,p_{v-1},\bar{p}_v)
\nonumber
\\
&+ \sum_{|\alpha| = \rho+1} 
\frac{(p_v-\bar{p}_v)^\alpha}{\rho!} \int_0^1  d\lambda 
\; (1-\lambda)^\rho \frac{\partial^{|\alpha|} F}{\partial p_v^\alpha}
(p_1,\dots,p_{v-1},\bar{p}_v+\lambda(p_v-\bar{p}_v)) \;,
\nonumber
\end{align}
where $\alpha$ is a multi-index. Together with the measure $dp_v$, the
last line combines with $\tanh^{-d}(\frac{\tau}{2})$ to a factor $dP
\;P^{\rho+1} \,e^{-P^2} \tanh^{\frac{\rho+1-d}{2}} (\frac{\tau}{2})$,
where $P=\frac{p_v-\bar{p}_v}{ \sqrt{\tanh \frac{\tau}{2}}}$.  For
sufficiently large but finite $\rho$ we shall see in (\ref{Gamma})
that the potential singularity in $t_0^{\frac{z}{2}}$ is cancelled so
that the last line of (\ref{Taylor}) is regular.
The bilinear form in the exponent has the form
\begin{align}
e^{-\frac{1}{4}pQ^{-1}p} &=
e^{-\frac{(p_v-\bar{p}_v)^2}{
4\omega\tanh \frac{\tau}{2}}
-\frac{1}{2}(p_v-\bar{p}_v)q -\frac{1}{2}\bar{p}_v
\mbox{\scriptsize$\displaystyle\sum_{j=1}^{v-1}$} \tilde{Q}_{vj}p_j
- \frac{1}{4}\mbox{\scriptsize$\displaystyle\sum_{i,j=1}^{v-1}$} 
\tilde{Q}_{ij} p_ip_j} \;,\quad
q :=\sum_{j=1}^{v-1} \tilde{Q}_{vj}p_j\;.
\end{align}
We can thus perform the
Gau\ss{}ian integration over $p_v$ and obtain for the restricted zeta
function $\zeta^r$ where the second line of (\ref{Taylor}) is removed:
\begin{align}
&\zeta^r_\phi(z) 
\label{zeta-Pv}
\\[-2ex]
&= \sum_{\mbox{\tiny$
\begin{array}{c}
k_1{+}2l_1{+}r_{11}{+}\dots{+} r_{1v}=n_1,\dots,\\
k_1{+}2l_1{+}r_{v1}{+}\dots{+} r_{vv}=n_v,\\
r_{ii}=0\;,~r_{ij}=r_{ji}
\end{array}$}}
\frac{n_1!\cdots n_v!}{\Gamma(\frac{z}{2})
\pi^{\frac{d}{2}}(2\sqrt{\pi})^{n_1+\dots+n_v}} 
\nonumber
\\
& \times \int_0^\infty dt_0 \; t_0^{\frac{z}{2}-1}
\int_0^\infty \prod_{i=1}^v\prod_{j_i=1}^{n_i} \frac{dt_{i,j_i} 
ds_{i,j_i}}{(t_{i,j_i}+s_{i,j_i})^{\frac{3}{2}}} \;
e^{-t} \Big(\frac{\omega}{
\tanh (\omega t)}\Big)^{\frac{d}{2}+\sum_{i=1}^v l_i+ \sum_{i<j} r_{ij}}
\nonumber
\\
& \times  
\int_{(\mathbb{R}^d)^{v-1}} 
\!\!\! \Big(\prod_{i=1}^{v-1} \frac{d p_i}{(2\pi)^d}\Big)
\sum_{|\alpha| \leq \rho}
\frac{(-2)^{|\alpha|}}{\alpha!} \frac{\partial^{|\alpha|}}{\partial q^\alpha}
\Big( 
e^{\frac{\omega}{4} q^2 \tanh(\omega t) -\frac{1}{2}\bar{p}_v
\sum_{j=1}^{v-1} \tilde{Q}_{vj}p_j
- \frac{1}{4}\sum_{i,j=1}^{v-1} \tilde{Q}_{ij} p_ip_j} \Big)
\nonumber
\\
& \times \frac{\partial^{|\alpha|}}{\partial p^\alpha_v} 
\Bigg( 
\bigg(\prod_{i < j}
\frac{\Big(\frac{2\cosh (\omega t_{ij})}{
\cosh (\omega t)} p_i p_j\Big)^{r_{ij}}}{r_{ij}!}\bigg)
\bigg(\prod_{i=1}^v 
\frac{\Big({\displaystyle\sum_{j\neq i}} \frac{\sinh (\omega t_{ij})}{
\sinh (\omega t)} p_i p_j\Big)^{k_i}}{k_i!}
\frac{(p_i^2)^{l_i}}{l_i!} \hat{f}_i(p_i)\bigg)\Bigg)_{p_v\mapsto
  \bar{p}_v}\;,
\nonumber
\end{align}
where
$t=\frac{1}{2\omega}\tau=t_0+\sum_{i=1}^v(T_i+S_i)$
and 
$t_{ij}=\frac{1}{2\omega}(\tau_{j}+\dots+\tau_{i-1})
-\frac{1}{2\omega}(\tau_{i}+\dots+\tau_{j-1})$.
The $q$-derivatives and the quadratic form in the exponent become 
with $\tilde{Q}_{ij}=\frac{\cosh(\omega t_{ij})-\cosh (\omega
  t)}{\omega \sinh (\omega t)}$
\begin{align}
\label{expand-q}
&\sum_{|\alpha| \leq \rho}
\frac{(-2)^{|\alpha|}}{\alpha!} \frac{\partial^{|\alpha|}}{\partial q^\alpha}
\Big( 
e^{\frac{\omega}{4} q^2 \tanh(\omega t) -\frac{1}{2}\bar{p}_v
\sum_{j=1}^{v-1} \tilde{Q}_{vj}p_j
- \frac{1}{4}\sum_{i,j=1}^{v-1} \tilde{Q}_{ij} p_ip_j} \Big)
\frac{\partial^{|\alpha|}}{\partial p^\alpha_v}
\\[-1ex]
& 
= \sum_{|\alpha|+2a \leq \rho}
(\omega \tanh (\omega t))^{a+|\alpha|} 
e^{-\big(\sum_{i,j=1}^{v-1}
\frac{
\sinh(\omega t^{-}_{ij})\sinh(\omega t^{+}_{ij})}{
2\omega \sinh(2\omega t)} p_ip_j\big)}
\nonumber
\\[-1ex]
&\qquad\quad \times \frac{1}{a!} 
\Big(\frac{\partial^2}{\partial p_v^\mu \partial p_{v\mu}}\Big)^{a}
\frac{1}{\alpha!} 
\bigg(\sum_{j=1}^{v-1} \frac{2 \sinh (\omega \frac{t+t_{vj}}{2})
\sinh (\omega \frac{t+t_{jv}}{2})}{\omega \sinh(\omega t)} p_j\bigg)^\alpha
\frac{\partial^{|\alpha|}}{\partial p^\alpha_v}\;,
\nonumber
\end{align}
where $t^{-}_{ij} = t+t_{kv}\big|_{k=\min(i,j)}$ and 
$t^{+}_{ij} = t+t_{vk}\big|_{k=\max(i,j)}$. Note that 
(\ref{expand-q}) is bounded for all $t$. 

We insert (\ref{expand-q}) into (\ref{zeta-Pv}).
We change the integration variables to
$t_0=(1-u)t$, $\sum_{i=1}^v (S_i+T_i)=ut$ with integration over $t$
from $0$ to $\infty$, over $u$ from $0$ to $1$ and over the surface
$\Delta$ given by $\sum_{i=1}^v (S_i+T_i)=1$. 
We write the denominators 
$\frac{1}{\sinh (\omega t)}= \frac{1}{(\omega t)} \cdot 
\frac{\omega t}{\sinh (\omega t)}$ and 
$\frac{1}{\tanh (\omega t)}= \frac{1}{(\omega t)} \cdot 
\frac{\omega t}{\tanh (\omega t)}$ 
and expand the bounded (at $0$)
fractions $\frac{\omega t}{\sinh (\omega t)}$ and 
$\frac{\omega t}{\tanh (\omega t)}$ into a Taylor 
series in $(\omega t)$. The numerators in hyperbolic functions of 
$(\omega t)$ and $(\omega t_{ij})$
and $\frac{1}{\cosh (\omega t)}$ are expanded into a Taylor series in their
arguments. Then, for each term in the sum,
the $u,t$-integral is of the form
\begin{align}
&\frac{1}{\Gamma(\frac{z}{2})}
\int_0^\infty dt
\;t^{(\frac{z}{2}-\frac{d}{2}+\frac{k_1+\dots+k_v}{2}+a+2|\alpha|+b-1)}    
e^{-t} \int_0^1 du (1-u)^{\frac{z}{2}-1} 
u^{\frac{n_1+\dots+n_v}{2}+c-1} 
\label{Gamma}
\\*
&= \frac{\Gamma(\tfrac{z}{2}-\tfrac{d}{2}+\tfrac{k_1+\dots+k_v}{2}
+a+2|\alpha|+b)\;\Gamma(\tfrac{n_1+\dots+n_v}{2}+c) 
}{\Gamma(\frac{z}{2}+\tfrac{n_1+\dots+n_v}{2}+c) }\;,
\nonumber
\end{align}
where the integers $b\geq c\geq 0$ arise from the Taylor expansion. 
The remaining integration over the simplex $\Delta$ is regular because from
the Taylor expansion only 
positive powers of the integration variables appear.
From (\ref{Gamma}) we deduce the following information 
about the pole structure:
\begin{itemize}
\item For $z\notin \mathbb{Z}$ or for $z > d$ there is no pole.

\item For $z=d-N$ with $N \in \mathbb{N}$, and $n_1,\dots,n_v$
  such that $z+n_1+\dots+n_v$ is even, there is a pole for a finite (and
  non-vanishing) number of index combinations and finite Taylor order
  $\rho=d+n_1+\dots+n_v-k_1-\dots-k_v$.
\end{itemize}
This concludes the proof that $\mathrm {Sd}=d-\mathbb{N}$.

It remains to characterise the nature of the residues. From
(\ref{zeta-Pv}) we conclude that the residues are given by the integral
over $p_1,\dots,p_{v-1}$ of an integrand which is a polynomial in
$p_1,\dots,p_{v-1}$ times $\prod_{i=1}^{v-1} \hat{f}_i(p_i)$ times
possible derivatives of $\hat{f}_v(\bar{p}_v)$. Reconstructing the
$p_v$-variable by a $\delta$-function and integrating by parts the
derivatives of $\hat{f}_v(\bar{p}_v)$, the residue becomes a finite
sum of the form 
\begin{align}
&\mathrm{res}_{z=d-N} (\zeta(z))
\\
&=\sum_{\alpha_0,\dots,\alpha_v} c_{\alpha_0\dots\alpha_v}
\int_{(\mathbb{R}^d)^{v}}   \Big(\prod_{i=1}^{v} \frac{d p_i}{(2\pi)^d}\Big)
\int_{\mathbb{R}^d} dx \;e^{\mathrm{i}(p_1+\dots+p_v)x} x^{\alpha_0}
\prod_{i=1}^{v} 
p_i^{\alpha_i}\hat{f}_i(p_i)
\nonumber
\\*
&=\sum_{\alpha_0,\dots,\alpha_v} \int_{\mathbb{R}^d}  dx 
\; c_{\alpha_0\dots\alpha_v} 
(-\mathrm{i})^{|\alpha_1|+\dots+|\alpha_v|} x^{\alpha_0}
\prod_{i=1}^{v} (\partial^{\alpha_i}f_i)(x_i)\;,
\nonumber
\end{align}
where the $\alpha_j$ are multi-indices which contract to a Lorentz
scalar. The prefactor $c_{\alpha_0\dots\alpha_v}$ results from the integration
over the $t$-variables. Thus, the residues are local. \hfill $\square$

\bigskip

We would like to stress that it was important to keep track of the
combinatorial factors which led to the cancellation of denominators 
$\frac{1}{\sinh \tau_i}$. Such denominators in the final formula
(\ref{zeta-Pv}) would be fatal because in that case the $u$-integral of
(\ref{Gamma}) would produce a hypergeometric function instead of the 
beta function and therefore an infinite sum for the residue, which
could be non-local. 

\section{Vertices contributing to the spectral action}

\label{App-B}

We compute here the individual vertex contributions (\ref{Duhamel}) to
the spectral action. This is done by inserting the vertices
(\ref{vertices}) into (\ref{Stf}) and computation of the
$t_i$-integrals.

\subsection{$V_2$}

The contribution of a single $V_2$-vertex is
\begin{align}
S_t(V_2)= \int_0^t dt_1 \;
\mathrm{tr}(e^{-\omega \Sigma t}) S^0_t(f)\;,\quad 
f=-2|\phi|^2-A_\mu A^\mu - B_\mu B^\mu\;.
\end{align}
With $\mathrm{tr}(e^{-\omega \Sigma t})=(2\cosh (\omega t))^4$ we have
after second order Taylor expansion, ignoring the remainder and the
odd first-order term,
\begin{align}
S_t(V_2) &= \int_{\mathbb{R}^4} \frac{dp}{(2\pi)^4} 
\frac{t}{(\tanh(\omega t))^4} 
\Big(\hat{f}(0) 
+ \frac{1}{2} p_\mu p_\nu \frac{\partial^2 \hat{f}}{\partial p_\mu 
\partial p_\nu}(0)\Big)
e^{-\frac{p^2}{4 \omega \tanh (\omega t)}}
\\
\nonumber
&= \frac{\omega^2 t}{\pi^2 \tanh^2(\omega t)} 
\Big(\hat{f}(0) 
+ \omega \tanh(\omega t) \delta_{\mu\nu} 
\frac{\partial^2 \hat{f}}{\partial p_\mu \partial p_\nu}(0)\Big)
\\
\nonumber
&= \frac{\omega^2 t}{\pi^2 \tanh^2(\omega t)} 
\int_{\mathbb{R}^4} dx \Big( f(x) 
- \omega \|x\|^2 \tanh(\omega t) \, f(x) \Big)\;,
\end{align}
after Fourier transformation $\hat{f}(p)=
\int_{\mathbb{R}^4} dx \;e^{-i p x} f(x)$.  
Inserting $f$ we obtain after Laplace transformation 
the leading terms of the asymptotic expansion to 
\begin{align}
\label{S2}
S_2(\mathcal{D}_A) &=  \frac{\chi_{-1}}{\pi^2}  
\int_{\mathbb{R}^4} dx  \;\big(-2|\phi|^2-A_\mu A^\mu-B_\mu
B^\mu\big)(x) 
\\*
\nonumber
& +  \frac{\chi_{0}}{\pi^2}  
\int_{\mathbb{R}^4} dx  \;(\omega^2 |x|^2 \big(2|\phi|^2+A_\mu A^\mu+B_\mu
B^\mu\big)(x)\;.
\end{align}

\subsection{$V_1 V_1$}

The contribution of two $V_1$-vertices is
\begin{align}
S_t(V_1,V_1)= \int_0^t dt_1 \int_0^{t-t_1} \!\! dt_2 \;
\mathrm{tr}(e^{-\omega \Sigma t}) S^{1,1}_{t_2,t-t_2}(-A,-A)
+ (A \mapsto B)\;.
\end{align}
This is the most involved computation.
To (\ref{Stf}) there are the two 
contributions $k_1=k_2=1$ 
up to order $0$ and $r_{12}=r_{21}=1$ 
with Taylor expansion about $p_2=-p_1$ up to order 2:

\begin{align}
&S_t(V_1,V_1) 
\\
&= \int_0^t dt_1 \int_0^{t-t_1} dt_2 
\int_{(\mathbb{R}^4)^2} \frac{dp_1 dp_2}{(2\pi)^8} 
\frac{1}{\tanh^4(\omega t)} \;
e^{-\frac{(p_1+p_2)^2}{4 \omega \tanh(\omega t) }
+ p_1p_2 \frac{\sinh(\omega t_2) \sinh(\omega(t-t_2))}{
\omega \sinh (\omega t)}}
\nonumber
\\
& \times 
\bigg\{
\hat{A}^{\mu}(p_1)\hat{A}^{\nu}(-p_1)
\Big(
\frac{\sinh^2 (\omega (t-2t_2))}{\sinh^2 (\omega t)} p_{1\mu} p_{1\nu}
+ 2\omega \delta_{\mu\nu}
\frac{\cos (\omega (t-2t_2))}{ \sinh (\omega t)} \Big)
\nonumber
\\
& + (p_1+p_2)^\rho \hat{A}^{\mu}(p_1)
\frac{\partial \hat{A}^{\nu}}{\partial p_2^\rho}(-p_1) \cdot 
2\omega \delta_{\mu\nu}
\frac{\cos (\omega (t-2t_2))}{ \sinh (\omega t)} 
\nonumber
\\
& + \frac{1}{2} (p_1+p_2)^\rho (p_1+p_2)^\sigma 
\hat{A}^{\mu}(p_1)\frac{\partial^2 \hat{A}^{\nu}}{\partial p_2^\rho 
\partial p_2^\sigma}(-p_1) \cdot 
2\omega \delta_{\mu\nu}
\frac{\cos (\omega (t-2t_2))}{ \sinh (\omega t)} \bigg\}
+ (A \mapsto B)
\nonumber
\\
&= \int_0^t dt_1 \int_0^{t-t_1} dt_2 
\int_{\mathbb{R}^4} \frac{dp_1}{(2\pi)^4} 
\frac{\omega^2}{\pi^2 \tanh^2(\omega t)} \;
e^{-\frac{\sinh(2\omega t_2) \sinh(2\omega(t-t_2))}{2\omega \sinh(2\omega t)}p_1^2 }
\nonumber
\\
& \times 
\bigg\{
\hat{A}^{\mu}(p_1)\hat{A}^{\nu}(-p_1)
\Big(
\frac{\sinh^2 (\omega (t-2t_2))}{\sinh^2 (\omega t)} p_{1\mu} p_{1\nu}
+ 2\omega \delta_{\mu\nu}
\frac{\cos (\omega (t-2t_2))}{ \sinh (\omega t)} \Big)
\nonumber
\\
& +4\omega p_1^\rho \hat{A}^{\mu}(p_1)
\frac{\partial \hat{A}_{\mu}}{\partial p_2^\rho }(-p_1) 
\cdot \frac{\sinh(\omega t_2)\sinh(\omega(t-t_2))}{ \cos (\omega t)}
\frac{\cos (\omega (t-2t_2))}{ \sinh (\omega t)} 
\nonumber
\\
& + \omega \Big(2 \delta^{\rho\sigma} \omega \tanh (\omega t) 
+ 4 p_1^\rho p_1^\sigma 
\frac{\sinh^2(\omega t_2) \sinh^2(\omega(t-t_2))}{ \cos^2 (\omega t)}\Big)
\nonumber
\\
& \qquad\qquad \times 
\hat{A}^{\mu}(p_1)\frac{\partial^2 \hat{A}_{\mu}}{\partial p_2^\rho 
\partial p_2^\sigma}(-p_1) \cdot 
\frac{\cos (\omega (t-2t_2))}{ \sinh (\omega t)} \bigg\}
+ (A \mapsto B)\;.
\nonumber
\end{align}
Up to $\mathcal{O}(t)$ this reduces to
\begin{align}
&S_t(V_1,V_1) 
\\
&= \int_0^t dt_1 \int_0^{t-t_1} dt_2 
\int_{\mathbb{R}^4} \frac{dp_1}{(2\pi)^4} 
\frac{\omega^2}{\pi^2 \tanh^2(\omega t)} \;
\nonumber
\\
& \times 
\bigg\{
\hat{A}^{\mu}(p_1)\hat{A}^{\nu}(-p_1)
\Big(
\frac{\sinh^2 (\omega (t-2t_2))}{\sinh^2 (\omega t)} p_{1\mu} p_{1\nu}
\nonumber
\\
& \qquad\qquad - \delta_{\mu\nu}
\frac{\cosh (\omega (t-2t_2))}{ \sinh (\omega t)} 
\frac{\sinh(2\omega t_2) \sinh(2\omega(t-t_2))}{\sinh(2\omega t)}p_1^2 
\Big)
\nonumber
\\
& 
+ 2\omega \hat{A}^{\mu}(p_1)\hat{A}_{\mu}(-p_1)
\frac{\cosh (\omega (t-2t_2))}{ \sinh (\omega t)} 
+ 2\omega^2 
\hat{A}^{\mu}(p_1)\frac{\partial^2 \hat{A}_{\mu}}{\partial p_2^\rho 
\partial p_{2,\rho}}(-p_1) 
\frac{\cosh (\omega (t{-}2t_2))}{ \cosh (\omega t)} \bigg\}
\nonumber
\\
&+ (A \mapsto B)
\nonumber
\\
&= \int_{\mathbb{R}^4} \frac{dp_1}{(2\pi)^4} 
\frac{\omega^2}{\pi^2 \tanh^2(\omega t)} \;
\nonumber
\\
& \times 
\bigg\{
\hat{A}^{\mu}(p_1)\hat{A}^{\nu}(-p_1)
\Big(
\Big( \frac{t}{4 \omega \tanh(\omega t)}
- \frac{t^2}{4 \sinh^2(\omega t)}\Big)p_{1\mu} p_{1\nu}
- \delta_{\mu\nu}
\frac{t \tanh(\omega t)}{6 \omega}p_1^2 
\Big)
\nonumber
\\
& 
+ t \, \hat{A}^{\mu}(p_1)\hat{A}_{\mu}(-p_1)
+ (\omega t) \tanh(\omega t)\,
\hat{A}^{\mu}(p_1)\frac{\partial^2 \hat{A}_{\mu}}{\partial p_2^\rho 
\partial p_{2\rho}}(-p_1) \bigg\}
+ (A \mapsto B)\;.
\nonumber
\end{align}
After Fourier and Laplace transformation, the leading contribution to
the spectral action becomes
\begin{align}
S_{11}(\mathcal{D}_A) &=  \frac{\chi_{-1}}{\pi^2}  
\int_{\mathbb{R}^4} dx  \;\big(A_\mu A^\mu+ B_\mu B^\mu\big)(x)  
\label{S11}
\\
&-  \frac{\chi_{0}}{\pi^2}  
\int_{\mathbb{R}^4} dx  \;(\omega^2 \|x\|^2) \big(A_\mu A^\mu+B_\mu B^\mu\big)(x)
\nonumber
\\
&- \frac{\chi_0}{12 \pi^2}
\int_{\mathbb{R}^4} d x \; \big(F^A_{\mu\nu} F^{A\mu\nu} + 
F^B_{\mu\nu} F^{B\mu\nu}\big)(x)\;.
\nonumber
\end{align}

\subsection{$V_2 V_2$, $V_3 V_3$, $V_4 V_4$ }

We have
\begin{align}
&\sum_{i=2}^4 S_t(V_i,V_i)
\\
&= \int_0^t dt_1 \int_0^{t-t_1} dt_2 
\bigg\{
\mathrm{tr}(e^{-\omega \Sigma t}) \Big( 
S^{0,0}_{t_2,t-t_2}(-|\phi|^2-A_\mu A^\mu,-|\phi|^2-A_\mu A^\mu)
\nonumber
\\
&\qquad\qquad\qquad 
+S^{0,0}_{t_2,t-t_2}(-|\phi|^2-B_\mu B^\mu,-|\phi|^2-B_\mu B^\mu)
\Big)
\nonumber
\\
&+\mathrm{tr}\Big(\frac{\mathrm{i}}{4}
[b^{\dag\mu}-b^\mu,b^{\dag\nu}-b^\nu]e^{-\omega \Sigma t_2}
\frac{\mathrm{i}}{4}
[b^{\dag\rho}-b^\rho,b^{\dag\sigma}-b^\sigma]e^{-\omega \Sigma (t-t_2)}
\Big) 
\nonumber
\\
&\qquad \times 
\Big( S^{0,0}_{t_2,t-t_2}(F^A_{\mu\nu},F^A_{\rho\sigma})
+S^{0,0}_{t_2,t-t_2}(F^B_{\mu\nu},F^B_{\rho\sigma})\Big)
\nonumber
\\
& + \mathrm{tr}\Big(
(b^{\dag\mu}-b^\mu)e^{-\omega \Sigma t_2}
(b^{\dag\nu}-b^\nu)e^{-\omega \Sigma (t-t_2)}
\Big) 
\nonumber
\\
&\qquad \times 
\Big( S^{0,0}_{t_2,t-t_2}(-D_\mu\phi ,\overline{D_\nu \phi})
+S^{0,0}_{t_2,t-t_2}(\overline{D_\mu \phi},-D_\nu\phi)
\Big)\bigg\}\;.
\nonumber
\end{align}
Since the $S^{0,0}_{t_2,t-t_2}$ are at least of $\mathcal{O}(t^{-2})$,
only the $\mathcal{O}(t^0)$-part of $e^{-\omega\Sigma t_2}$ and
$e^{-\omega\Sigma (t-t_2)}$ will contribute to the spectral
action. Now the traces in $\bigwedge (\mathbb{C}^4)$ are easy 
to compute:
\begin{align}
\mathrm{tr}\big(e^{ \omega \Sigma t}\big) &=(2\cosh (\omega t))^4\;, 
\\
\mathrm{tr}\big((b^{\dag\mu} -b^\mu)e^{ -\omega \Sigma t_2} 
(b^{\dag\nu} -b^\nu)e^{ -\omega \Sigma (t-t_2)} \big) 
&=-16 \delta^{\mu\nu} + \mathcal{O}(t)\;, 
\nonumber
\\
\mathrm{tr}\big(\tfrac{\mathrm{i}}{4}[b^{\dag\mu} {-}b^\mu,
b^{\dag\nu} {-}b^\nu]e^{ -\omega \Sigma t_2} 
\tfrac{\mathrm{i}}{4}[b^{\dag\rho} {-}b^\rho,
b^{\dag\sigma} {-}b^\sigma]e^{ -\omega \Sigma (t-t_2)} \big)  
&= 8 (\delta^{\mu\rho}\delta^{\nu\sigma}
{-} \delta^{\mu\sigma}\delta^{\nu\rho}) 
{+} \mathcal{O}(t).
\nonumber
\end{align}
After Taylor expansion about $p_2=-p_1$ up to order $0$, 
integration over $p_2,t_1,t_2$ and Laplace transformation 
we obtain 
\begin{align}
\label{S22}
(S_{22}{+}S_{33}{+}S_{44})(\mathcal{D}_A) = \frac{\chi_0}{2\pi^2} &
\int_{\mathbb{R}^4} \!\! d x \Big\{ 2\overline{D_\mu \phi} (D^\mu \phi)
+(|\phi|^2+A_\mu A^\mu)^2
\\
& 
+F_{\mu\nu}^A F^{A\mu\nu}
+ (|\phi|^2+B_\mu B^\mu)^2 
+F_{\mu\nu}^B F^{B\mu\nu}
\Big\}(x)\;.
\nonumber
\end{align}

\subsection{$V_1 V_2$, $V_2 V_1$}

With the abbreviation $f_{\phi A}:=|\phi|^2+ A_\mu A^\mu$,
we have
\begin{align}
&S_t(V_1,V_2)+S_t(V_2,V_1)
\\
&= \int_0^t dt_1 \int_0^{t-t_1} dt_2 \;
\mathrm{tr}(e^{-\omega \Sigma t}) 
\Big( S^{1,0}_{t_2,t-t_2}(-A,-f_{\phi A})
+S^{0,1}_{t_2,t-t_2}(-f_{\phi A},-A) \Big) 
\nonumber
\\
&+ (A \mapsto B)
\nonumber
\\
&=\int_0^t dt_1 \int_0^{t-t_1} dt_2 \; 
\int_{\mathbb{R}^4 \times \mathbb{R}^4} 
\frac{dp_1 dp_2}{(2\pi)^8} 
\frac{1}{\tanh^4(\omega t)} 
\nonumber
\\
& \times \Big(
p_{2,\mu}\hat{A}^{\mu}(p_1)\hat{f}_{\phi A}(p_2) 
-p_{1,\mu}\hat{A}^{\mu}(p_2)\hat{f}_{\phi A}(p_1) \Big)
\frac{\sinh (\omega (t-2t_2))}{\sinh (\omega t)} 
e^{-\frac{1}{4} pQ^{-1}p} + (A \mapsto B)
\nonumber
\\
&=\int_0^t dt_1 \int_0^{t-t_1} dt_2 \; 
\int_{\mathbb{R}^4}
\frac{dp_1}{(2\pi)^4} 
\frac{\omega^2}{\pi^2 \tanh^2(\omega t)} 
\frac{\sinh (\omega (t-2t_2))}{\sinh (\omega t)} 
\nonumber
\\
& \times \Big(
-p_{1,\mu}\hat{A}^{\mu}(p_1)\hat{f}_{\phi A}(-p_1) 
-p_{1,\mu}\hat{A}^{\mu}(-p_1)\hat{f}_{\phi A}(p_1) \Big)
+ (A \mapsto B) + \mathcal{O}(t)
\nonumber
\\
&= \mathcal{O}(t) \;.\nonumber
\end{align}
We thus have 
\begin{align}
S_{12}(\mathcal{D}_A)=0\;.
\label{S12}
\end{align}

\subsection{$V_1V_1 V_2$,  $V_1V_2 V_1$,  $V_2V_1 V_1$} 

Only the $k_i=0$ terms in (\ref{Stf}) contribute to the leading
order. 
With the abbreviation $f_{\phi A}:=|\phi|^2+ A_\mu A^\mu$, these give
\begin{align}
&S_t(V_1,V_1,V_2)+S_t(V_1,V_2,V_1)+S_t(V_2,V_1,V_1)
\\
&= \int_0^t dt_1 \int_0^{t-t_1} dt_2 \int_0^{t-t_1-t_2} dt_3 \;
\mathrm{tr}(e^{-\omega \Sigma t}) 
\Big( S^{1,1,0}_{t_3,t_2,t-t_2-t_3}(-A,-A,-f_{\phi A})
\nonumber
\\
&\qquad
+S^{1,0,1}_{t_3,t_2,t-t_2-t_3}(-A,-f_{\phi A},-A) 
+S^{0,1,1}_{t_3,t_2,t-t_2-t_3}(-f_{\phi A},-A,-A) 
\Big) + (A \mapsto B)
\nonumber
\\
&= \int_0^t dt_1 \int_0^{t-t_1} dt_2 \int_0^{t-t_1-t_2} dt_3 \;
\int_{(\mathbb{R}^4)^3}\frac{dp_1\,dp_2\,dp_3}{(2\pi)^{12}} 
\frac{-2\omega}{\tanh^4(\omega t) \sinh(\omega t)}
\nonumber
\\
&\times \Big(
\hat{A}_\mu(p_1) \hat{A}^\mu(p_2) \hat{f}_{\phi A}(p_3) \cosh(\omega(t-2t_3))
\nonumber
\\
& \qquad
+\hat{A}_\mu(p_1) \hat{f}_{\phi A}(p_2) \hat{A}^\mu(p_3)  
\cosh(\omega(t-2t_2-2t_3))
\nonumber
\\
& \qquad
+\hat{f}_{\phi A}(p_1) \hat{A}_\mu(p_2) \hat{A}^\mu(p_3)  
\cosh(\omega(t-2t_2))
\Big) e^{-\frac{1}{4} pQ^{-1}p}  + (A \mapsto B) + \mathcal{O}(t)
\nonumber
\\
&= \int_0^t dt_1 \int_0^{t-t_1} dt_2 \int_0^{t-t_1-t_2} dt_3 \;
\int_{(\mathbb{R}^4)^2}\frac{dp_1\,dp_2}{(2\pi)^{8}} 
\frac{(-2\omega^3) }{\pi^2 \tanh^2(\omega t) \sinh(\omega t)}
\nonumber
\\
&\times 
\hat{A}_\mu(p_1) \hat{A}^\mu(p_2) \hat{f}_{\phi A}(-p_1-p_2) 
\Big(\cosh(\omega(t-2t_3))
+\cosh(\omega(t-2t_2-2t_3))
\nonumber
\\
& \qquad\qquad\qquad +\cosh(\omega(t-2t_2))
\Big) + (A \mapsto B)+ \mathcal{O}(t)
\nonumber
\\
&= \int_{(\mathbb{R}^4)^2}\frac{dp_1\,dp_2}{(2\pi)^{8}} 
\frac{(-\omega^2 t^2) }{\pi^2 \tanh^2(\omega t)}
\hat{A}_\mu(p_1) \hat{A}^\mu(p_2) \hat{f}_{\phi A}(-p_1-p_2) 
+(A\mapsto B) + \mathcal{O}(t)\;.
\nonumber
\end{align}
After Fourier and Laplace transformation we obtain
\begin{align}
S_{112}(\mathcal{D}_A) = -\frac{\chi_0}{\pi^2} &
\int_{\mathbb{R}^4} \!\! d x \Big\{
A_\mu A^\mu(|\phi|^2+A_\nu A^\nu )
+B_\mu B^\mu(|\phi|^2+B_\nu B^\nu)\Big\}(x)\;.
\label{S112}
\end{align}

\subsection{$V_1 V_1 V_1$}

The leading order in (\ref{Stf}) is given by the 
$(k_1=1,r_{23}=1)$ and the other two cyclic permutations:
\begin{align}
&S_t(V_1,V_1,V_1)
\\
&= \int_0^t dt_1 \int_0^{t-t_1} dt_2 \int_0^{t-t_1-t_2} dt_3 \;
\mathrm{tr}(e^{-\omega \Sigma t}) S^{1,1,1}_{t_3,t_2,t-t_2-t_3}(-A,-A,-A)
+ (A \mapsto B)
\nonumber
\\
&= \int_0^t \!\!\! dt_1 \int_0^{t-t_1} \!\!\!\!\!\! dt_2 \int_0^{t-t_1-t_2} 
\!\!\!\!\!\! dt_3 \int_{(\mathbb{R}^4)^3} \!\!\!\!\! 
\frac{dp_1\,dp_2\,dp_3}{(2\pi)^{12}} 
\frac{-2\omega}{\tanh^4(\omega t) \sinh^2(\omega t)}
\hat{A}_\mu(p_1) \hat{A}_\nu(p_2) \hat{A}_\rho(p_3) 
\nonumber
\\
&\times 
\Big(
(p_2^\mu \sinh (\omega(t-2t_3)) + p_3^\mu \sinh (\omega(t-2t_2-2t_3)))
\delta^{\nu\rho} \cosh (\omega(t-2t_2)) 
\nonumber
\\*
&\quad 
+ (p_3^\nu \sinh (\omega(t-2t_2)) + p_1^\nu \sinh (\omega (2t_3-t)) )
\delta^{\rho\mu} \cosh (\omega(t-2t_2-2t_3))
\nonumber
\\*
&\quad 
+ (p_1^\rho \sinh (\omega(2t_2 + 2 t_3-t)) +p_2^\rho \sinh (\omega (2t_2-t)))
\delta^{\mu\nu} \cosh (\omega(t-2t_3)) \Big)
\nonumber
\\
& \times 
e^{-\frac{1}{4}pQ^{-1}p} +(A \mapsto B)
\nonumber
\\
&= \int_0^t \!\!\! dt_1 \int_0^{t-t_1} \!\!\!\!\!\! dt_2 \int_0^{t-t_1-t_2} 
\!\!\!\!\!\! dt_3 \int_{(\mathbb{R}^4)^3} \!\!\!\!\! 
\frac{dp_1\,dp_2\,dp_3}{(2\pi)^{12}} 
\frac{-2\omega}{\tanh^4(\omega t) \sinh^2(\omega t)}
\hat{A}_\mu(p_1) \hat{A}_\nu(p_2) \hat{A}_\rho(p_3) 
\nonumber
\\
&\times p_2^\mu \Big(
\sinh (\omega(2t_2-2t_3))
+ \sinh (\omega(4t_3+2t_2-2t)
+\sinh (\omega(2t -4 t_2-2t_3))
\Big)
\nonumber
\\
&+(A \mapsto B)
\nonumber
\\
&= \mathcal{O}(t)\;.
\nonumber
\end{align}
(The integral without $e^{-\frac{1}{4}pQ^{-1}p}$ cancels exactly.) We thus have
\begin{align}
S_{111}(\mathcal{D}_A)=0\;.
\label{S111}
\end{align}

\subsection{$V_1 V_1 V_1 V_1$}

The leading order in (\ref{Stf}) is given by the three possibilities with 
$k_i=0$:
\begin{align}
&S_t(V_1,V_1,V_1,V_1)
\\
&= \int_0^t dt_1 \int_0^{t-t_1} dt_2 \int_0^{t-t_1-t_2} dt_3 
\int_0^{t-t_1-t_2-t_3} dt_4 \;
\mathrm{tr}(e^{-\omega \Sigma t}) 
\nonumber
\\
& \qquad\qquad \times 
S^{1,1,1,1}_{t_4,t_3,t_2,t-t_2-t_3-t_4}(-A,-A,-A,-A)
+ (A \mapsto B)
\nonumber
\\
&= 
\int_{(\mathbb{R}^4)^3} \!\!\!\!\! 
\frac{dp_1\,dp_2\,dp_3}{(2\pi)^{12}} \;
\frac{(2\omega)^2}{\tanh^4(\omega t) \sinh^2(\omega t)}
\hat{A}_\mu(p_1) \hat{A}_\nu(p_2) \hat{A}_\rho(p_3) \hat{A}_\sigma(p_4) 
\nonumber
\\
&\times 
\int_0^t \!\!\! dt_1 \int_0^{t-t_1} \!\!\!\!\!\! dt_2 
\int_0^{t-t_1-t_2} \!\!\!\!\!\! dt_3 
\int_0^{t-t_1-t_2} \!\!\!\!\!\! dt_3 
\int_0^{t-t_1-t_2-t_3} \!\!\!\!\!\! dt_4 
\Big(
\cosh (\omega t_{21}) \cosh (\omega t_{43})
\delta^{\mu\nu} \delta^{\rho\sigma} 
\nonumber
\\
&\qquad +\cosh (\omega t_{31}) \cosh (\omega t_{42})
\delta^{\mu\rho} \delta^{\nu\sigma} 
+\cosh (\omega t_{41}) \cosh (\omega t_{32})
\delta^{\mu\sigma} \delta^{\nu\rho} 
\Big)+ (A\mapsto B)\;,
\nonumber
\end{align}
with $t_{21}=t-2t_4$, $t_{43}=t-2t_2$, $t_{31}=t-2t_3-2t_4$,
$t_{42}=t-2t_2-2t_3$, $t_{41}=t-2t_2-2t_3-2t_4$ and $t_{32}=t-2t_3$.
Taylor expansion in $p_4$ and Gau\ss{}ian integration over
$\frac{dp_4}{(2\pi)^4}$ yield, as usual, a factor
$\frac{\omega^2}{\pi^2} \tanh(\omega t)$ and an exponential function
which can be ignored in leading order. The $t_1\dots t_4$ integrals
evaluate to $\frac{t^2 \sinh^2(\omega t)}{8\omega^2}$, so that we
conclude
\begin{align}
S_{1111}(\mathcal{D}_A)= \frac{\chi_0}{2\pi^2}
\int_{\mathrm{R}^4}dx\;\Big\{A_\mu A^\mu A_\nu A^\nu + 
B_\mu B^\mu B_\nu B^\nu \Big\}(x)\;.
\label{S1111}
\end{align}

\end{appendix}

\end{document}